\documentclass[twocolumn]{aastex7} %
\usepackage[utf8]{inputenc}
\usepackage{hyperref}
\usepackage{float}
\usepackage{graphicx}
\usepackage{wrapfig}
\usepackage{tablefootnote}
\usepackage{threeparttable}
\usepackage{subfigure}
\usepackage{natbib}
\usepackage{subcaption}
\usepackage{hhline}
\usepackage{amsmath}
\usepackage{amssymb}

\begin{document}

\newcommand{\Cornell}{Department of Astronomy and Cornell Center for Astrophysics and Planetary Science, Cornell University, Ithaca, NY, 14853, USA}
\newcommand{\UCB}{Department of Astronomy, University of California, Berkeley, 501 Campbell Hall \#3411, Berkeley, CA 94720, USA}

\correspondingauthor{Ben Jacobson-Bell}
\email{ben.jacobson-bell@nanograv.org}

\author[0009-0009-6231-9280]{Ben Jacobson-Bell}
\email{}
\affiliation{\UCB}
\affiliation{\Cornell}

\author[0000-0002-4049-1882]{James M. Cordes}
\email{}
\affiliation{\Cornell}

\author[0000-0002-2878-1502]{Shami Chatterjee}
\email{}
\affiliation{\Cornell}

\author[0009-0006-5007-7470]{Sashabaw Niedbalski}
\email{}
\affiliation{\Cornell}

\author[0000-0001-5134-3925]{Gabriella Agazie}
\affiliation{Center for Gravitation, Cosmology and Astrophysics, Department of Physics and Astronomy, University of Wisconsin-Milwaukee,\\ P.O. Box 413, Milwaukee, WI 53201, USA}
\email{}
\author[0000-0002-8935-9882]{Akash Anumarlapudi}
\affiliation{Department of Physics and Astronomy, University of North Carolina, Chapel Hill, NC 27599, USA}
\email{}
\author[0000-0003-0638-3340]{Anne M. Archibald}
\affiliation{Newcastle University, NE1 7RU, UK}
\email{}
\author[0009-0008-6187-8753]{Zaven Arzoumanian}
\affiliation{X-Ray Astrophysics Laboratory, NASA Goddard Space Flight Center, Code 662, Greenbelt, MD 20771, USA}
\email{}
\author[0000-0002-4972-1525]{Jeremy G. Baier}
\affiliation{Department of Physics, Oregon State University, Corvallis, OR 97331, USA}
\email{}
\author[0000-0003-2745-753X]{Paul T. Baker}
\affiliation{Department of Physics and Astronomy, Widener University, One University Place, Chester, PA 19013, USA}
\email{}
\author[0000-0003-3053-6538]{Paul R. Brook}
\affiliation{Institute for Gravitational Wave Astronomy and School of Physics and Astronomy, University of Birmingham, Edgbaston, Birmingham B15 2TT, UK}
\email{}
\author[0000-0002-6039-692X]{H. Thankful Cromartie}
\affiliation{National Research Council Research Associate, National Academy of Sciences, Washington, DC 20001, USA resident at Naval Research Laboratory, Washington, DC 20375, USA}
\email{}
\author[0000-0002-1529-5169]{Kathryn Crowter}
\affiliation{Department of Physics and Astronomy, University of British Columbia, 6224 Agricultural Road, Vancouver, BC V6T 1Z1, Canada}
\email{}
\author[0000-0002-2185-1790]{Megan E. DeCesar}
\altaffiliation{Resident at the Naval Research Laboratory}
\affiliation{Department of Physics and Astronomy, George Mason University, Fairfax, VA 22030, resident at the U.S. Naval Research Laboratory, Washington, DC 20375, USA}
\email{}
\author[0000-0002-6664-965X]{Paul B. Demorest}
\affiliation{National Radio Astronomy Observatory, 1003 Lopezville Rd., Socorro, NM 87801, USA}
\email{}
\author[0000-0002-2554-0674]{Lankeswar Dey}
\affiliation{Department of Physics and Astronomy, West Virginia University, P.O. Box 6315, Morgantown, WV 26506, USA}
\affiliation{Center for Gravitational Waves and Cosmology, West Virginia University, Chestnut Ridge Research Building, Morgantown, WV 26505, USA}
\email{}
\author[0000-0001-8885-6388]{Timothy Dolch}
\affiliation{Department of Physics, Hillsdale College, 33 E. College Street, Hillsdale, MI 49242, USA}
\affiliation{Eureka Scientific, 2452 Delmer Street, Suite 100, Oakland, CA 94602-3017, USA}
\email{}
\author[0000-0001-7828-7708]{Elizabeth C. Ferrara}
\affiliation{Department of Astronomy, University of Maryland, College Park, MD 20742, USA}
\affiliation{Center for Research and Exploration in Space Science and Technology, NASA/GSFC, Greenbelt, MD 20771}
\affiliation{NASA Goddard Space Flight Center, Greenbelt, MD 20771, USA}
\email{}
\author[0000-0001-5645-5336]{William Fiore}
\affiliation{Department of Physics and Astronomy, University of British Columbia, 6224 Agricultural Road, Vancouver, BC V6T 1Z1, Canada}
\email{}
\author[0000-0001-8384-5049]{Emmanuel Fonseca}
\affiliation{Department of Physics and Astronomy, West Virginia University, P.O. Box 6315, Morgantown, WV 26506, USA}
\affiliation{Center for Gravitational Waves and Cosmology, West Virginia University, Chestnut Ridge Research Building, Morgantown, WV 26505, USA}
\email{}
\author[0000-0001-7624-4616]{Gabriel E. Freedman}
\affiliation{NASA Goddard Space Flight Center, Greenbelt, MD 20771, USA}
\email{}
\author[0000-0001-6166-9646]{Nate Garver-Daniels}
\affiliation{Department of Physics and Astronomy, West Virginia University, P.O. Box 6315, Morgantown, WV 26506, USA}
\affiliation{Center for Gravitational Waves and Cosmology, West Virginia University, Chestnut Ridge Research Building, Morgantown, WV 26505, USA}
\email{}
\author[0000-0001-8158-683X]{Peter A. Gentile}
\affiliation{Department of Physics and Astronomy, West Virginia University, P.O. Box 6315, Morgantown, WV 26506, USA}
\affiliation{Center for Gravitational Waves and Cosmology, West Virginia University, Chestnut Ridge Research Building, Morgantown, WV 26505, USA}
\email{}
\author[0000-0003-4090-9780]{Joseph Glaser}
\affiliation{Department of Physics and Astronomy, West Virginia University, P.O. Box 6315, Morgantown, WV 26506, USA}
\affiliation{Center for Gravitational Waves and Cosmology, West Virginia University, Chestnut Ridge Research Building, Morgantown, WV 26505, USA}
\email{}
\author[0000-0003-1884-348X]{Deborah C. Good}
\affiliation{Department of Physics and Astronomy, University of Montana, 32 Campus Drive, Missoula, MT 59812}
\email{}
\author[0000-0003-2742-3321]{Jeffrey S. Hazboun}
\affiliation{Department of Physics, Oregon State University, Corvallis, OR 97331, USA}
\email{}
\author[0000-0003-1082-2342]{Ross J. Jennings}
\altaffiliation{NANOGrav Physics Frontiers Center Postdoctoral Fellow}
\affiliation{Department of Physics and Astronomy, West Virginia University, P.O. Box 6315, Morgantown, WV 26506, USA}
\affiliation{Center for Gravitational Waves and Cosmology, West Virginia University, Chestnut Ridge Research Building, Morgantown, WV 26505, USA}
\email{}
\author[0000-0001-6607-3710]{Megan L. Jones}
\affiliation{Center for Gravitation, Cosmology and Astrophysics, Department of Physics and Astronomy, University of Wisconsin-Milwaukee,\\ P.O. Box 413, Milwaukee, WI 53201, USA}
\email{}
\author[0000-0001-6295-2881]{David L. Kaplan}
\affiliation{Center for Gravitation, Cosmology and Astrophysics, Department of Physics and Astronomy, University of Wisconsin-Milwaukee,\\ P.O. Box 413, Milwaukee, WI 53201, USA}
\email{}
\author[0000-0002-0893-4073]{Matthew Kerr}
\affiliation{Space Science Division, Naval Research Laboratory, Washington, DC 20375-5352, USA}
\email{}
\author[0000-0003-0721-651X]{Michael T. Lam}
\affiliation{SETI Institute, 339 N Bernardo Ave Suite 200, Mountain View, CA 94043, USA}
\affiliation{School of Physics and Astronomy, Rochester Institute of Technology, Rochester, NY 14623, USA}
\affiliation{Laboratory for Multiwavelength Astrophysics, Rochester Institute of Technology, Rochester, NY 14623, USA}
\email{}
\author[0000-0001-6436-8216]{Bjorn Larsen}
\affiliation{Department of Physics, Yale University, New Haven, CT 06511, USA}
\email{}
\author[0000-0003-1301-966X]{Duncan R. Lorimer}
\affiliation{Department of Physics and Astronomy, West Virginia University, P.O. Box 6315, Morgantown, WV 26506, USA}
\affiliation{Center for Gravitational Waves and Cosmology, West Virginia University, Chestnut Ridge Research Building, Morgantown, WV 26505, USA}
\email{}
\author[0009-0006-9938-157X]{Georgia A. Lowes}
\affiliation{E.A. Milne Centre for Astrophysics, University of Hull, Cottingham Road, Kingston-upon-Hull, HU6 7RX, UK}
\affiliation{Centre of Excellence for Data Science, Artificial Intelligence and Modelling (DAIM), University of Hull, Cottingham Road, Kingston-upon-Hull, HU6 7RX, UK}
\email{}
\author[0000-0001-5373-5914]{Jing Luo}
\altaffiliation{Deceased}
\affiliation{Department of Astronomy \& Astrophysics, University of Toronto, 50 Saint George Street, Toronto, ON M5S 3H4, Canada}
\email{}
\author[0000-0001-5229-7430]{Ryan S. Lynch}
\affiliation{Green Bank Observatory, P.O. Box 2, Green Bank, WV 24944, USA}
\email{}
\author[0000-0001-8313-0895]{Ashley Martsen}
\affiliation{Department of Physics and Astronomy, West Virginia University, P.O. Box 6315, Morgantown, WV 26506, USA}
\affiliation{Center for Gravitational Waves and Cosmology, West Virginia University, Chestnut Ridge Research Building, Morgantown, WV 26505, USA}
\email{}
\author[0000-0001-5481-7559]{Alexander McEwen}
\affiliation{Center for Gravitation, Cosmology and Astrophysics, Department of Physics and Astronomy, University of Wisconsin-Milwaukee,\\ P.O. Box 413, Milwaukee, WI 53201, USA}
\email{}
\author[0000-0001-7697-7422]{Maura A. McLaughlin}
\affiliation{Department of Physics and Astronomy, West Virginia University, P.O. Box 6315, Morgantown, WV 26506, USA}
\affiliation{Center for Gravitational Waves and Cosmology, West Virginia University, Chestnut Ridge Research Building, Morgantown, WV 26505, USA}
\email{}
\author[0000-0002-4642-1260]{Natasha McMann}
\affiliation{Department of Physics and Astronomy, Vanderbilt University, 2301 Vanderbilt Place, Nashville, TN 37235, USA}
\email{}
\author[0000-0001-8845-1225]{Bradley W. Meyers}
\affiliation{Australian SKA Regional Centre (AusSRC), Curtin University, Bentley, WA 6102, Australia}
\affiliation{International Centre for Radio Astronomy Research (ICRAR), Curtin University, Bentley, WA 6102, Australia}
\email{}
\author[0000-0002-2689-0190]{Patrick M. Meyers}
\affiliation{ETH Zurich, Institute for Particle Physics and Astrophysics, Wolfgang-Pauli-Strasse 27, 8093 Zurich, Switzerland}
\email{}
\author[0000-0002-3616-5160]{Cherry Ng}
\affiliation{Dunlap Institute for Astronomy and Astrophysics, University of Toronto, 50 St. George St., Toronto, ON M5S 3H4, Canada}
\email{}
\author[0000-0002-0940-6563]{Mason Ng}
\affiliation{Department of Physics, McGill University, 3600  University St., Montreal, QC H3A 2T8, Canada}
\affiliation{Trottier Space Institute at McGill University, 3550 rue University, Montr\'{e}al, QC H3A 2A7, Canada}
\email{}
\author[0000-0002-6709-2566]{David J. Nice}
\affiliation{Department of Physics, Lafayette College, Easton, PA 18042, USA}
\email{}
\author[0009-0001-1750-3531]{Shania Nichols}
\altaffiliation{NANOGrav Physics Frontiers Center Postdoctoral Fellow}
\affiliation{SETI Institute, 339 N Bernardo Ave Suite 200, Mountain View, CA 94043, USA}
\email{}
\author[0000-0002-7374-6925]{Daniel J. Oliver}
\altaffiliation{NANOGrav Physics Frontiers Center Postdoctoral Fellow}
\affiliation{Department of Physics, Oregon State University, Corvallis, OR 97331, USA}
\email{}
\author[0000-0001-5465-2889]{Timothy T. Pennucci}
\affiliation{Institute of Physics and Astronomy, E\"{o}tv\"{o}s Lor\'{a}nd University, P\'{a}zm\'{a}ny P. s. 1/A, 1117 Budapest, Hungary}
\email{}
\author[0000-0002-8509-5947]{Benetge B. P. Perera}
\affiliation{Arecibo Observatory, HC3 Box 53995, Arecibo, PR 00612, USA}
\email{}
\author[0000-0002-8826-1285]{Nihan S. Pol}
\affiliation{Department of Physics, Texas Tech University, Box 41051, Lubbock, TX 79409, USA}
\email{}
\author[0000-0002-2074-4360]{Henri A. Radovan}
\affiliation{Department of Physics, University of Puerto Rico, Mayag\"{u}ez, PR 00681, USA}
\email{}
\author[0000-0001-5799-9714]{Scott M. Ransom}
\affiliation{National Radio Astronomy Observatory, 520 Edgemont Road, Charlottesville, VA 22903, USA}
\email{}
\author[0000-0002-5297-5278]{Paul S. Ray}
\affiliation{Space Science Division, Naval Research Laboratory, Washington, DC 20375-5352, USA}
\email{}
\author[0000-0001-7832-9066]{Alexander Saffer}
\altaffiliation{NANOGrav Physics Frontiers Center Postdoctoral Fellow}
\affiliation{National Radio Astronomy Observatory, 520 Edgemont Road, Charlottesville, VA 22903, USA}
\email{}
\author[0000-0003-4391-936X]{Ann Schmiedekamp}
\affiliation{Department of Physics, Penn State Abington, Abington, PA 19001, USA}
\email{}
\author[0000-0002-1283-2184]{Carl Schmiedekamp}
\affiliation{Department of Physics, Penn State Abington, Abington, PA 19001, USA}
\email{}
\author[0000-0002-7283-1124]{Brent J. Shapiro-Albert}
\affiliation{Department of Physics and Astronomy, West Virginia University, P.O. Box 6315, Morgantown, WV 26506, USA}
\affiliation{Center for Gravitational Waves and Cosmology, West Virginia University, Chestnut Ridge Research Building, Morgantown, WV 26505, USA}
\affiliation{Giant Army, 915A 17th Ave, Seattle WA 98122}
\email{}
\author[0000-0001-9784-8670]{Ingrid H. Stairs}
\affiliation{Department of Physics and Astronomy, University of British Columbia, 6224 Agricultural Road, Vancouver, BC V6T 1Z1, Canada}
\email{}
\author[0000-0002-7261-594X]{Kevin Stovall}
\affiliation{National Radio Astronomy Observatory, 1003 Lopezville Rd., Socorro, NM 87801, USA}
\email{}
\author[0000-0002-2820-0931]{Abhimanyu Susobhanan}
\affiliation{Max-Planck-Institut f{\"u}r Gravitationsphysik (Albert-Einstein-Institut), Callinstra{\ss}e 38, D-30167 Hannover, Germany\\Leibniz Universit{\"a}t Hannover, D-30167 Hannover, Germany}
\email{}
\author[0000-0002-1075-3837]{Joseph K. Swiggum}
\altaffiliation{NANOGrav Physics Frontiers Center Postdoctoral Fellow}
\affiliation{Department of Physics, Lafayette College, Easton, PA 18042, USA}
\email{}
\author[0009-0001-5938-5000]{Mercedes S. Thompson}
\affiliation{Department of Physics and Astronomy, University of British Columbia, 6224 Agricultural Road, Vancouver, BC V6T 1Z1, Canada}
\email{}
\author[0009-0002-6412-7812]{Amir Tresnjic}
\affiliation{ETH Zurich, Institute for Particle Physics and Astrophysics, Wolfgang-Pauli-Strasse 27, 8093 Zurich, Switzerland}
\email{}
\author[0000-0001-9678-0299]{Haley M. Wahl}
\affiliation{Department of Physics and Astronomy, West Virginia University, P.O. Box 6315, Morgantown, WV 26506, USA}
\affiliation{Center for Gravitational Waves and Cosmology, West Virginia University, Chestnut Ridge Research Building, Morgantown, WV 26505, USA}
\email{}


\title{The NANOGrav 15\,yr and 20\,yr Datasets: Timing Events and Pulse Shape Changes}

\begin{abstract}
The average pulse shape of a pulsar is typically stable over decadal timescales, enabling estimation of pulse times of arrival to better than a small fraction of the pulse width  using matched filtering techniques. However, in North American Nanohertz Observatory for Gravitational Waves (NANOGrav) observations of PSR\,J1713+0747, three discrete timing events that depart from the prevailing timing model have been seen in the last 20\,yr. All three correspond to morphological changes in pulse shape. Using principal component analysis, we analyze the pulse profiles of nine NANOGrav pulsars, including seven with profiles from the 15\,yr dataset and two with additional profiles from the forthcoming 20\,yr dataset. We recover the three known pulse shape change events in PSR\,J1713+0747 and another previously known event in PSR\,J1643$-$1224. We implement a ranking metric for candidate events and address four highly ranked candidates in this nine-pulsar sample. We also recover known slow pulse shape variations in PSR\,J1643$-$1224, PSR\,J1903+0327, and PSR\,B1937+21 and report an unexpected recurrence after $\sim$10\,yr of one such variation in PSR\,B1937+21.
\end{abstract}

\keywords{radio astronomy --- millisecond pulsars --- interstellar medium}

\section{Introduction}\label{sec:intro}

The pulses emitted by radio pulsars have characteristic average shapes that are typically stable on long ($\gtrsim$10\,yr) timescales. When $10^3$--$10^5$ single pulses are averaged at a specific epoch to approximate this stable shape, the product is called a pulse profile. Pulsar timing for the detection of nanohertz-scale gravitational waves, first suggested by \cite{Sazhin78} and \citet{Detweiler79}, leverages the stability of pulse profiles to estimate minuscule deviations from their expected times of arrival (TOAs). Such deviations, if correlated between all observed pulsars across the sky in agreement with the Hellings--Downs curve \citep{HellingsDowns83}, point toward the existence of a stochastic nanohertz gravitational-wave background (GWB). However, this analysis requires that pulse TOAs can be estimated to extremely high ($<$1\,$\mu$s) precision, so much work has focused on modeling the noise processes that contaminate pulsar timing data and obscure the GWB signal.

Astrophysical sources of nonstationary noise in pulsar timing include stochastic spin noise, timing glitches, emission state changes, and pulse shape change events, among other phenomena. Interpretation of these departures from idealized timing models invariably boils down to the question of whether they are due to extrinsic propagation effects or intrinsic to the pulsar, either internal to the neutron star or from its magnetosphere.   

Pulsars spin down due to the loss of their rotational kinetic energy as magnetic dipole radiation; glitches are sudden, rapid ``spin-up'' events. Glitches may accompany profile shape changes \citep{Keith25}.
State changes include discrete jumps between a small number of profile shapes \citep[``mode changes'';][]{Bartel82}, discrete jumps between a small number of quantized subpulse drift rates, and switching between high- and low-intensity states. Collectively, state changes are consistent with a first-order Markov process in many cases \citep{Cordes13}. They may also correspond to changes in torque, as demonstrated for ``intermittent'' pulsars with state changes on $\sim$1\,month and longer time scales \citep{Kramer06, Lyne10}.

We define ``pulse shape change events'' as a class of profile changes that occur suddenly ($<$30\,d) before the pulse shape gradually relaxes back to something approximating its previous morphology (though subtle changes can persist over long timescales). Pulsar timing requires the precise fitting of pulse TOAs to a model informed by the pulse period, the spin-down rate, the pulse shape, and other variables. Sudden deviations from these variables' expected values can result in ``timing events,'' which limit the effectiveness of this method for detecting a GWB, but such deviations can be mitigated if they can be adequately incorporated in the timing model.

Pulsar timing arrays (PTAs) such as the North American Nanohertz Observatory for Gravitational Waves (NANOGrav) are built preferentially from millisecond pulsars (MSPs), named for their unusually short rotational periods and famous for the high stability of their pulse shapes relative to canonical pulsars. The NANOGrav PTA consisted of 68 MSPs at the time of the 15\,yr data release \citep{NG15ObsTiming}, all of which were chosen for their pulse stability and/or brightness. No glitches or mode switching events have been observed in any NANOGrav pulsar to date. However, four total pulse shape change events have been observed in two NANOGrav pulsars, resulting in significant noise increases in each pulsar's timing contribution.

Three of those events were observed in the pulse profiles of PSR J1713+0747. These events are chromatic, occurring to a greater degree at lower frequencies, but do not appear to follow the $\nu^{-2}$ dependence expected if their chromaticity were due to changes in electron density along the line of sight \citep{Jennings24}. \citet{Demorest13} and \citet{Lam18} report the first and second events, respectively, as chromatic timing events. \citet{Xu21} report the pulse shape change for the third event, while \citet{Jennings24} provide a detailed analysis of the correlation between the shape change and the TOA and DM residuals. It was only recognized after the detection of the third event, in analyses by \citet{Lin21} and \citet{Goncharov21}, that all three timing events corresponded to pulse shape change events.

Prior to the detection of the 2021 pulse shape change event in PSR J1713+0747, a similar event was observed by \citet{Shannon16} in PSR J1643$-$1224: a new pulse component became visible before slowly decaying away, though not without leaving the pulse profile minutely but permanently altered. Curiously, \citet{Shannon16} find that the effect of the PSR J1643$-$1224 event on TOA residuals is greatest around 10\,cm (3\,GHz), weaker at 20\,cm (1.5\,GHz), and essentially absent at 50\,cm (600\,MHz), while \citet{Jennings24} show that the 2021 PSR J1713+0747 event is strongest at 800\,MHz and weaker at both 600\,MHz and 1.5\,GHz.

\citet{Goncharov21} report a pulse shape change event in a third NANOGrav pulsar, PSR\,0437$-$4715. The event occurred in early 2015, nearly coincident with the \citet{Shannon16} event in PSR\,J1643$-$1224. However, NANOGrav did not begin timing PSR\,0437$-$4715 until the second half of 2015, so the event does not appear in the 15\,yr dataset and we defer its analysis to other work.

In this work, we address timing difficulties caused by pulse profile variability, focusing on sudden, unforeseeable pulse shape change events like those described above.  We introduce a detection method for such events using principal component analysis (PCA) and apply it to nine NANOGrav pulsars. We find four event candidates at a level higher than at least one of those already known. We attribute two to long-term pulse shape variability in PSR\,B1937+21, previously analyzed by \citet{Brook18} and interpreted as scattering variability. Of the other two, both low-S/N, one matches a set of epochs excised from NANOGrav timing analysis due to anomalous DMX values, or deviations from the pulsar's average dispersion measure (DM), and the other appears to also affect its pulsar's interpulse. In addition to the one-off event candidates found by our search algorithm, we observe nonstationary features in the time series of PCA coefficients for several pulsars, including but not limited to the scattering variability in PSR\,B1937+21, demonstrating that PCA is an effective tool for pulse profile variability analysis over a range of timescales. 

\subsection{Data Sample}

We study nine NANOGrav pulsars: PSR\,J0030+0451, PSR\,J1022+1001, PSR\,J1600$-$3053, PSR\,J1643$-$1224, PSR\,J1713+0747, PSR\,J1903+0327, PSR\,J1909$-$3744, PSR\,B1937+21, and PSR\,J2234+0611, listed in order by right ascension. 

PSR\,J1643$-$1224 and PSR\,J1713+0747 are known to have undergone pulse shape change events at a combined four epochs \citep{Shannon16, Lin21}. The existence of these events in NANOGrav observations is the primary motivation for this work and the recovery of the events is an important criterion for our search.

PSR J1909$-$3744 and PSR J2234+0611 are, respectively, the two least noisy pulsars in the NANOGrav PTA, where by ``least noisy'' we mean the definition used in the leave-one-out analysis of \citet{NG15OneByOne}: the lowest weighted rms of TOA residuals after epoch averaging, timing model fitting, and red noise removal. Their data are not expected to contain significant pulse shape change events; we include them as a baseline against which to compare other pulsars in our sample.

We study PSR\,J1903+0327 because it is known to exhibit long-term timing effects from exceptionally strong interstellar scattering along the line of sight \citep{Geiger25} and PSR\,J1022+1001 because it has been shown by \citet{Fiore25} to undergo significant intra-epoch pulse shape variation, potentially contributing to noticeable epoch-to-epoch variation. We selected the remaining two pulsars in our sample, PSR\,J0030+0451 and PSR\,J1600$-$3053, by visual inspection of their 15\,yr timing and DM residuals. To the eye, these pulsars appear to exhibit occasional DM discontinuities consistent with those seen for known pulse shape change events.

A version of this analysis implemented over the full NANOGrav PTA would place limits on the overall prevalence of pulse shape change events in NANOGrav timing data. However, many pulsars in the PTA have not been observed for long enough to compare putative events, which may last hundreds or thousands of days, to a stable-shape baseline. We defer the implementation of this analysis on the full PTA to future work.

\section{Observations}\label{sec:obs}

NANOGrav observations in the 15\,yr dataset \citep{NG15ObsTiming} took place at three observatories: the Green Bank Observatory (GBO), the Arecibo Observatory (AO), and the Karl G. Jansky Very Large Array (VLA). Since the Arecibo Telescope's collapse in 2020 December, observations have also been made with the Canadian Hydrogen Intensity Mapping Experiment (CHIME); however, these were not part of the 15\,yr dataset, which concludes with the Arecibo Telescope's collapse. We use only 15\,yr data in this work, except for two pulsars in our sample: PSR\,J1713+0747, for which we use GBO observations from the same time interval as was analyzed by \citet{Jennings24}, which extends through the end of 2022; and PSR\,B1937+21, for which we use GBO data up to 2021.5. In the latter case, this is to show the recurrence of a slow pulse shape change event $\sim$10\,yr after it first occurred.

In this work, we use GBO observations for five of our pulsars and AO observations for the other four. NANOGrav has used the VLA since 2015 to observe five pulsars, all of them in our sample, at 3\,GHz \citep{NG15ObsTiming}, but we defer the anaysis of VLA data to future work since, for now, they occupy a relatively short fraction of the overall observing period. 

The 15\,yr dataset includes both GBO and AO data for two of our pulsars, PSR\,J1713+0747 and PSR\,B1937+21, which coincidentally are the two pulsars for which we extend our analysis beyond the 15\,yr dataset. These observations were taken with similar frequency coverage ($\sim$1--2\,GHz) to control for instrumental errors across the PTA. For our analysis, we prioritize the GBO data for continuity beyond the 15\,yr dataset. In Section \ref{sec:B1937+21}, we compare GBO and AO data for PSR\,B1937+21 during the two slow events it underwent, though the events appear most strongly in GBO 800\,MHz data and our instrumental control analysis is limited by the only overlapping frequencies being above 1\,GHz.

GBO observations use the 100\,m Robert C. Byrd Green Bank Telescope (GBT). Three back ends have been used for NANOGrav observations on the GBT: the Green Bank Astronomical Signal Processor \citep[GASP;][]{DemorestASP} from 2006 to 2011, the Green Bank Ultimate Pulsar Processing Instrument \citep[GUPPI;][]{DuPlainGUPPI} from 2010 to 2020, and the Versatile GBT Astronomical Spectrometer (VEGAS) since 2019. These intervals overlap due to trial periods when new back ends were implemented. VEGAS observations are not part of the 15\,yr dataset, but are included in this work as part of the extended datasets for PSR\,J1713+0747 and PSR\,B1937+21.

As shown in the Table 1 of \citet{NG15ObsTiming}, GBT data processed by the GUPPI back end span 722--919\,MHz when using the 800\,MHz receiver and 1151--1885\,MHz when using the $L$-band receiver. VEGAS data have frequency ranges approximately equivalent to GUPPI data. GASP data span a narrower range: 822--866\,MHz for the lower-frequency receiver and 1386--1434\,MHz for the higher.

At AO, NANOGrav observations used the Arecibo Signal Processor \citep[ASP;][]{DemorestASP} and the Puerto Rican Ultimate Pulsar Processing Instrument (PUPPI) to digitize data from the 430\,MHz, $L$-band, and $S$-band receivers. (The 327\,MHz receiver has also been used, but only for one pulsar and not one in our sample.) The ASP and PUPPI back ends covered bandwidths comparable to GASP and GUPPI, respectively.

The NANOGrav preprocessing pipeline, which involves calibration, radio-frequency interference (RFI) removal, normalization, and reduction, ends in .ff data products, which are the pulse profiles we use for our search. The process is described for the 15\,yr dataset in the Section 3.1 of \citet{NG15ObsTiming} and is similar for the 20\,yr dataset. After this stage, some epochs are excised from the timing analysis for various reasons described in the Appendix A of \citet{NG15ObsTiming}. Of these, we retain only the cut due to a malfunctioning local oscillator at AO between MJD 57984 and MJD 58447. The cut for eclipsing pulsars is irrelevant for our sample. The other cuts target outliers, often resulting from instrumental errors and typically isolated to a few epochs at most. We leave in most of these epochs to compare our method's performance on real pulse shape change events versus one-off outliers.

We excise one epoch, MJD\,55977, from GBO data on the basis of instrumental error. \citet{Brook18} comment that on this day of observations with the GBT, only a noise diode (and not an on-sky source) was used for calibration, and that the effect on pulse shape was notable. The same diode-only calibration scheme was used on other days, too, but did not affect the pulse shape of the observed pulsars as drastically as on MJD\,55977, so we leave them in for our analysis. We excise a separate epoch, MJD\,55135, from AO data for PSR\,J1903+0327 on the basis of atypical contamination in the off-pulse region.

\section{Methods}\label{sec:methods}

\subsection{Principal Component Analysis}\label{sec:pca}

\begin{figure*}
    \centering
\includegraphics[width=\textwidth]{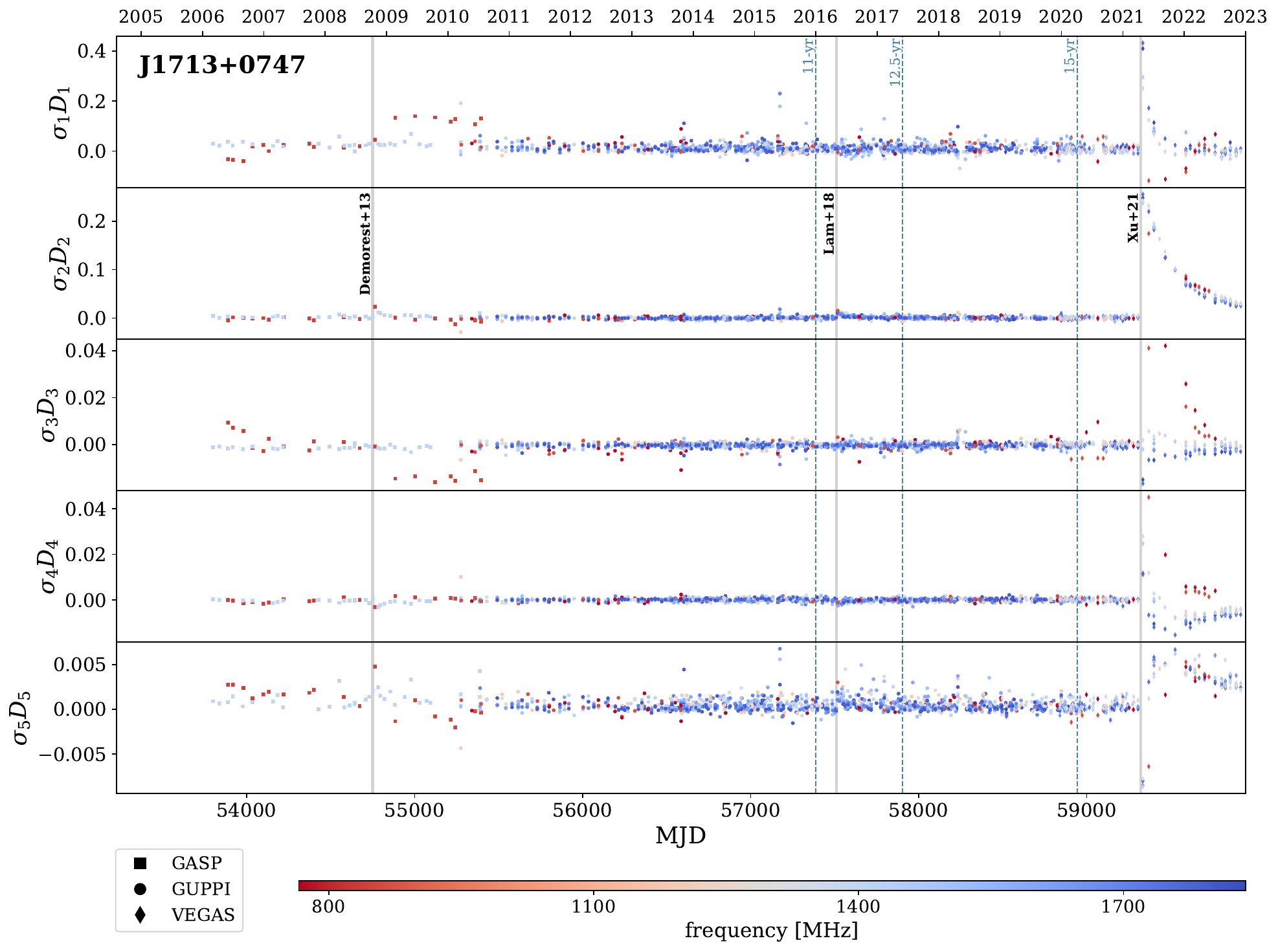}
    \caption{Time series of projection magnitudes (that is, dot products, $D_i$) of PSR\,J1713+0747 pulse profile residuals ($\delta U$) along each of the first five PCs of the dataset. The dot products are normalized by the square root of each PC's eigenvalue, $\sigma_i$, for comparison across PCs. The vertical gray lines give the epochs reported by \citet{Demorest13}, \citet{Lam18}, and \citet{Xu21} for their respective timing events. The 2021 event is clearly visible at MJD\,59320 in all five panels. The 2008 and 2016 events are also visible in some panels. The vertical blue dashed lines give the ends of the time ranges in the NANOGrav 11\,yr, 12.5\,yr, and 15\,yr data releases. This is a figure set; images for all pulsars in our sample, including their interpulses, are available in the online journal (11 images). }
    \label{fig:pca}
\end{figure*}

\begin{figure}
    \centering
    \includegraphics[width=0.49\textwidth]{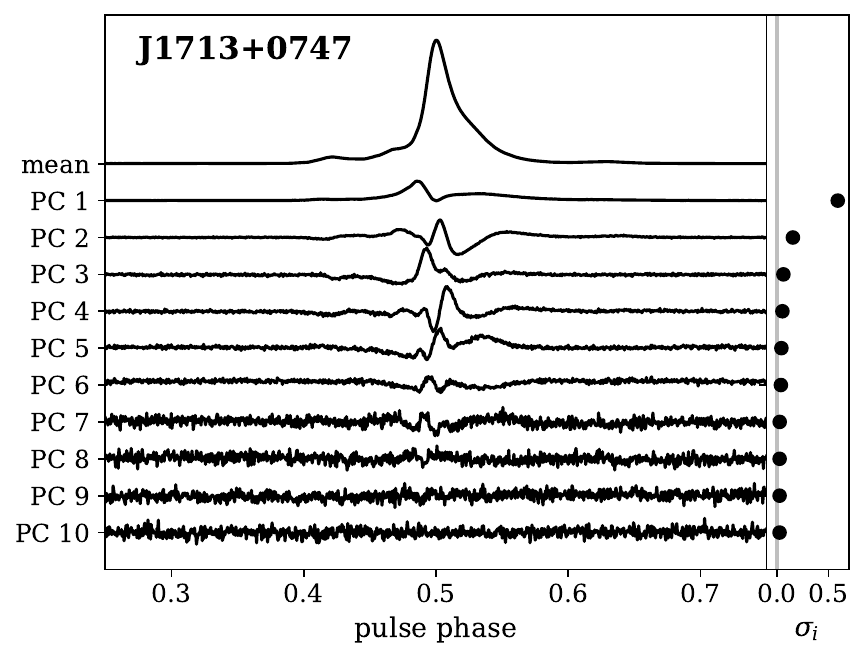}
    \caption{Left panel: the mean pulse profile for PSR\,J1713+0747 in our dataset, followed by the first ten PCs. Elements are zeroed outside the central 50\% of pulse phase. Successive PCs capture decreasing amounts of the variance in the dataset, with the result that after the first several PCs, they begin to resemble white noise. 
    Right panel: the square roots of the eigenvalues corresponding to each of the first ten PCs. This is a figure set; images for all pulsars in our sample, including their interpulses, are available in the online journal (11 images). 
    }
    \label{fig:comps}
\end{figure}

\begin{figure}
    \centering
    \includegraphics[width=\columnwidth]{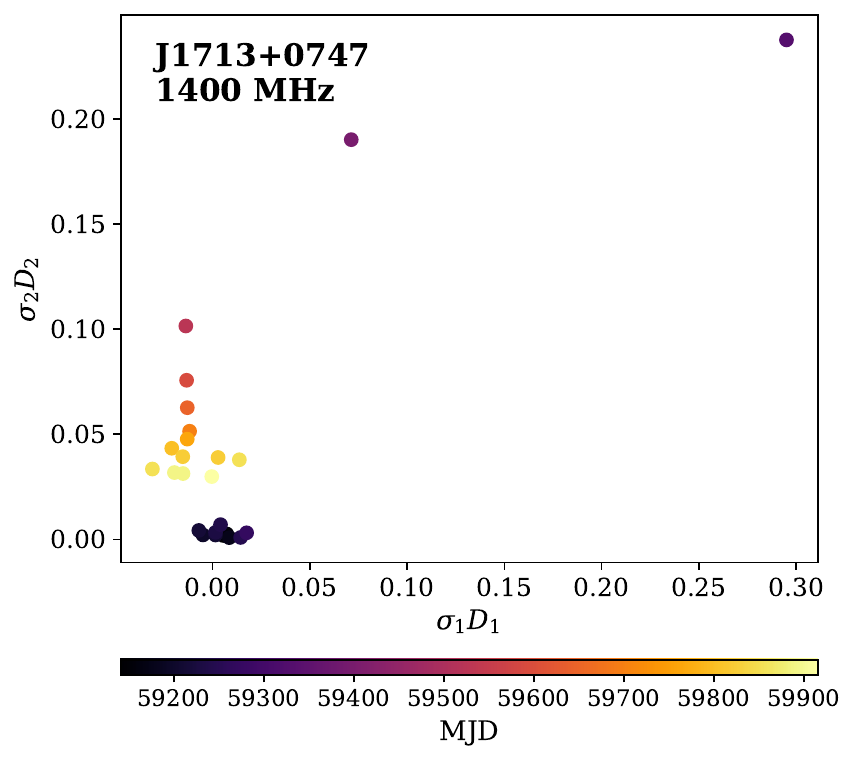}
    \caption{
    Scatter plot of the dot products, $D_i$, of PSR J1713+0747 pulse profile residuals ($\delta U$) with their first two PCs, scaled by $\sigma_i$. For clarity, only the $D_i$ of 1400\,MHz profiles are shown. Color denotes the epoch of observation, beginning 200\,d before the 2021 event. A discontinuity is visible around MJD\,59320; the points then trace a curve back toward the locus of pre-event points, but do not reach it.}
    \label{fig:scatter:J1713+0747}
\end{figure}

\begin{center}
\begin{table}
\begin{threeparttable}
    \caption{Pulse Profile S/N Cutoffs and Profile Counts for PC Calculation}
    \begin{tabular}{c c c}
        \hline
        \hline
        Pulsar & S/N cutoff & $N_\text{profiles}$ \\
        \hline
        J1713+0747 & 50 & 1519 \\
        B1937+21 & 50 & 1351 \\
        J1643$-$1224 & 20 & 1201 \\
        J1600$-$3053 & 20 & 1055 \\
        J0030+0451 & 20 & 875 \\
        J1909$-$3744 & 50 & 800 \\
        J1903+0327 & 10 & 586 \\
        J1022+1001 & 20 & 509 \\
        J2234+0611 & 15 & 257 \\
        \hline
    \end{tabular}
    \label{tab:snrcuts}
\end{threeparttable}
\end{table}
\end{center}

PCA is a method for studying the morphological variability of a dataset. It decomposes a dataset of similar vectors (pulse profiles, in this case) into a basis of eigenvectors (``principal components,'' or PCs) ordered by the statistical variance they represent in the dataset. In pulse profile analysis, PCA has previously been used by \citet{Jennings24:PCA} to study the effects of pulse jitter on timing. \citet{Lin21} used PCA in their study of the first two pulse shape change events (then called ``timing events,'' as they had been seen in time series of TOA residuals before the associated pulse shape changes were confirmed) in PSR\,J1713+0747. \citet{Jennings24} used PCA to study the third pulse shape change event in this pulsar, confirming its existence in GBT data at both 800 and 1500\,MHz and in data from CHIME. 
Throughout this section, we use PSR\,J1713+0747 as a case study for our method, requiring that our method recover the three known events in this pulsar and the one known event in PSR\,J1643--1224 while minimizing its response to single-day outliers, which are likely to be instrumental. Our results for PSR\,J1713+0747 are summarized in Section \ref{sec:J1713+0747} and for PSR\,J1643--1224 in Section \ref{sec:J1643-1224}, while complete details for the other seven pulsars in our sample are given in the remainder of Section \ref{sec:results}.

We preprocess our data by aligning each profile using
template matching as implemented in Fourier space
 \citep{Taylor92}\footnote{Implemented in PyPulse \citep{PyPulse}.} and normalizing to unit amplitude. This process requires a pulse template, an idealized pulse shape generated from the average of years of observations for each pulsar. After aligning, we zero out bins outside the central 50\% of pulse phase, a preprocessing step also used by \citet{Lin18}. Two pulsars in our sample have an interpulse; in each of these cases, we run the analysis twice, treating the main pulse and interpulse separately. In this paper's figures and tables, a lowercase ``i'' is appended to the pulsar's name (e.g., B1937+21i) if the figure refers to its interpulse.

We use the peak flux divided by the rms of off-pulse phase bins to estimate the S/N for each profile and impose an S/N cutoff below which profiles are thrown out. The off-pulse window is manually defined for each pulsar, no fewer than 300 (out of 2048) phase bins, after an inspection of the pulsar's template, which ensures we do not capture any component of the main pulse or interpulse in the window.
Since different pulsars in our sample have different levels of noise, the S/N cutoff is also set manually for each pulsar and is chosen by inspection of the S/N distribution. For consistency, cutoff values of 50 and 20 are preferred unless neither would result in a sufficiently large sample of profiles ($\gtrsim$250) for PCA, in which case lower values are considered. A list of cutoffs is given in Table \ref{tab:snrcuts}.

Rather than taking a single average pulse profile across all frequencies for each epoch, we average the spectrogram in 100\,MHz subbands to explore the chromaticity of pulse shape change events. For GUPPI and VEGAS observations, this gives us seven pulse profiles per observation from the $L$-band receiver and two from the 800\,MHz receiver. For PUPPI observations, where the usable bandwidth was narrower than the back end processing bandwidth due to sensitivity loss at the edges of the bandpass, we obtain six pulse profiles per observation from the $L$-band receiver and one from the $430$\,MHz receiver. (For ASP and GASP, the processing band was much narrower; we obtain only one subaveraged profile apiece for each receiver.) The chromaticity of pulse shape change events in PCA was previously studied by \citet{Jennings24} for receiver-to-receiver variations. While \citet{Jennings24} showed the chromaticity of TOA residuals using 25\,MHz bins, however, the chromaticity of pulse shape variation using PCA has not been reported on frequency scales finer than a receiver bandwidth. We define our subaverage bins starting from the first-indexed frequency within the appropriate sensitivity range of each observation, which may vary epoch to epoch by $\lesssim2$\,MHz. Accordingly, small variations occur in the center frequencies of our subaverage bins.

In the alignment and normalization process, we match profiles from each subband to a unique template, taking into account chromatic variations in the pulse shape. These variations occur to some extent in all pulsars, but should not be confused with the rare shape change events we seek, so we do not let them influence the PCA. Our wideband ``pulse portraits,'' a 2D analog to pulse templates, are due to \citet{Pennucci19}, are handled with PulsePortraiture \citep{PulsePortraiture}, and are available for download as part of the NANOGrav 15\,yr dataset \citep{NG15ObsTiming, NG15data}. To emphasize departures from the expected pulse shape, we use the same frequency-dependent pulse portrait for all observations of a given pulsar; we do not modify the template to better fit any apparent pulse shape changes.

Since the 2021 PSR\,J1713+0747 event exhibits chromaticity inconsistent with a $\nu^{-2}$ power law, \citet{Jennings24} and others have concluded that pulse shape change events broadly are not a DM effect. \citet{Shannon16} ascribe the PSR\,J1643$-$1224 event, which also does not follow a $\nu^{-2}$ trend, to a disturbance of the pulsar's magnetosphere; that is, an effect local to the pulsar and not the line of sight. NANOGrav .ff data products are dedispersed to a fiducial DM for each pulsar, but DM may vary by epoch by $\sim$0.001--0.01\,pc\,cm$^{-3}$, which can affect the shape of a frequency-averaged pulse profile. Prior to subaveraging, we align all channels with PyPulse \citep{PyPulse}, which mitigates this effect.

PCA involves the computation of the covariance matrix of a zero-centered version of a dataset (that is, one in which each vector has had the ensemble's average vector subtracted off). In our framework, the vectors are 2048-element pulse profiles, so the average vector is the average of many pulse profiles and looks very like the standard pulse template. The covariance matrix's eigenvectors are the PCs, ranked in descending order by their eigenvalues, where each eigenvalue gives the variance in the system captured by the corresponding eigenvector. The eigensystem is typically solved by singular value decomposition. \citet{Lin18} give a thorough overview of the process; we use the \verb+scikit-learn+ \citep{scikit-learn} implementation.

Projecting any profile $U$ along the direction of the $i$th PC (that is, taking the dot product of the two vectors, $U\cdot \text{PC }i$) gives PC $i$'s contribution to the deviation of profile $U$ from the average, which can be a useful quantity for comparing pulse shapes across profiles. However, pulse shapes have a known stable frequency dependence, and projecting the pulse profiles along PCs leaves in this effect. By instead projecting the profile residuals $\delta U$ along PCs, $D_i \equiv \delta U \cdot \text{PC }i$, where each residual is produced by the subtraction of a frequency-dependent template from the pulse profile, we remove this chromaticity.

For each pulsar in our sample, we study the time series of dot products of the pulse profile residuals $\delta U$ with each of the first five PCs. For these time series, we impose the same S/N cutoff as was used to calculate the PCs, which is given for each pulsar in Table \ref{tab:snrcuts} alongside the number of profiles represented in the PC calculation and time series. Figure \ref{fig:pca} is a figure set giving the $D_i$ time series for each pulsar in our sample, including their interpulses if they exist. In the second panel of the figure for PSR\,J1713+0747, note the coincidence of the black dashed lines, denoting the reported epochs of the three known pulse shape change events, with discontinuities and exponential decays in the time series. The first ten PCs and the square roots of their eigenvalues are shown in Figure \ref{fig:comps}, also a figure set with additional figures in the online journal showing the PCs for the other pulsars in our sample.

It can also be illustrative to view multiple PC projections at once in a scatter plot; one for the 1400\,MHz subband of the first two PCs is given in Figure \ref{fig:scatter:J1713+0747}. In this space, the 2021 pulse shape change event is clearly visible as a jump from the locus of black circles near (0,0) to the far upper right of the plot. As the color brightens with increasing time, the points move back toward the locus at the bottom; the pulse shape relaxes back to approximately what it was before the event, though small differences remain.

\subsection{Pulse Shape Change Event Search}

\begin{figure}
    \centering
    \includegraphics[width=\columnwidth]{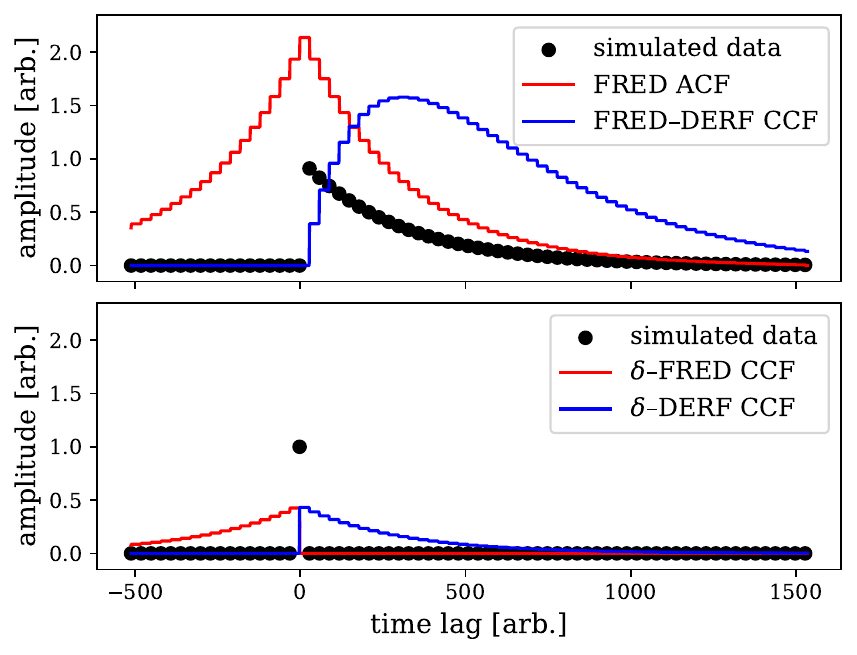}
    \caption{Toy model of our matched filtering framework on two basic time series--like vectors of length 2048. The top panel shows an evenly sampled ($\delta t = 30$) FRED event with decay time $\tau = 300$; the bottom panel shows a single-pixel spike (a delta function) with the same sampling. Both time series are convolved with a FRED template and a DERF template, both with $\tau = 300$, and the CCFs overplotted in the corresponding panels. The ratio of the FRED CCF peak amplitude to the DERF CCF peak amplitude is $e/2$ for the FRED event and 1 for the delta function. The time lag between the FRED and DERF peaks is equal to $\tau$ for the FRED event and equal to $\delta t$ for the delta function.}
    \label{fig:matchfiltertoy}
\end{figure}

\begin{figure}
    \centering
    \includegraphics[width=\columnwidth]{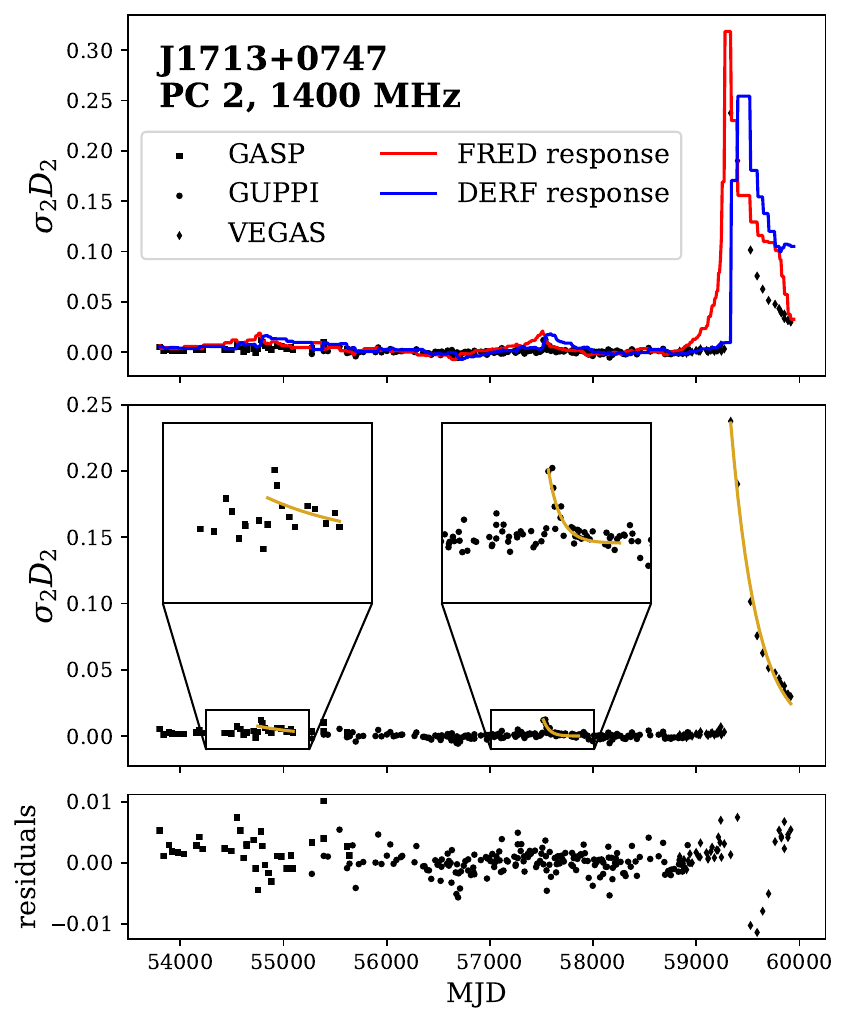}
    \caption{Top panel: Application of our matched filtering framework to the $D_2$ time series for PSR\,J1713+0747 in the 1400\,MHz subband. For demonstration, we use $\tau = 100$\,d for our template decay time. 
    Middle panel: Best-fit decaying exponentials to an interval of $D_2$ following the epoch of each event. The interval is 1\,yr for the first two events and $\sim$1.7\,yr for the third, since the event is clearly still occurring by the end of our dataset.
    Bottom panel: Residuals for the exponential fits from the middle panel. Systematic underestimates followed by overestimates are visible for the third event.}
    \label{fig:matchfilterex}
\end{figure}

\begin{figure*}
    \centering
    \includegraphics[width=\textwidth]{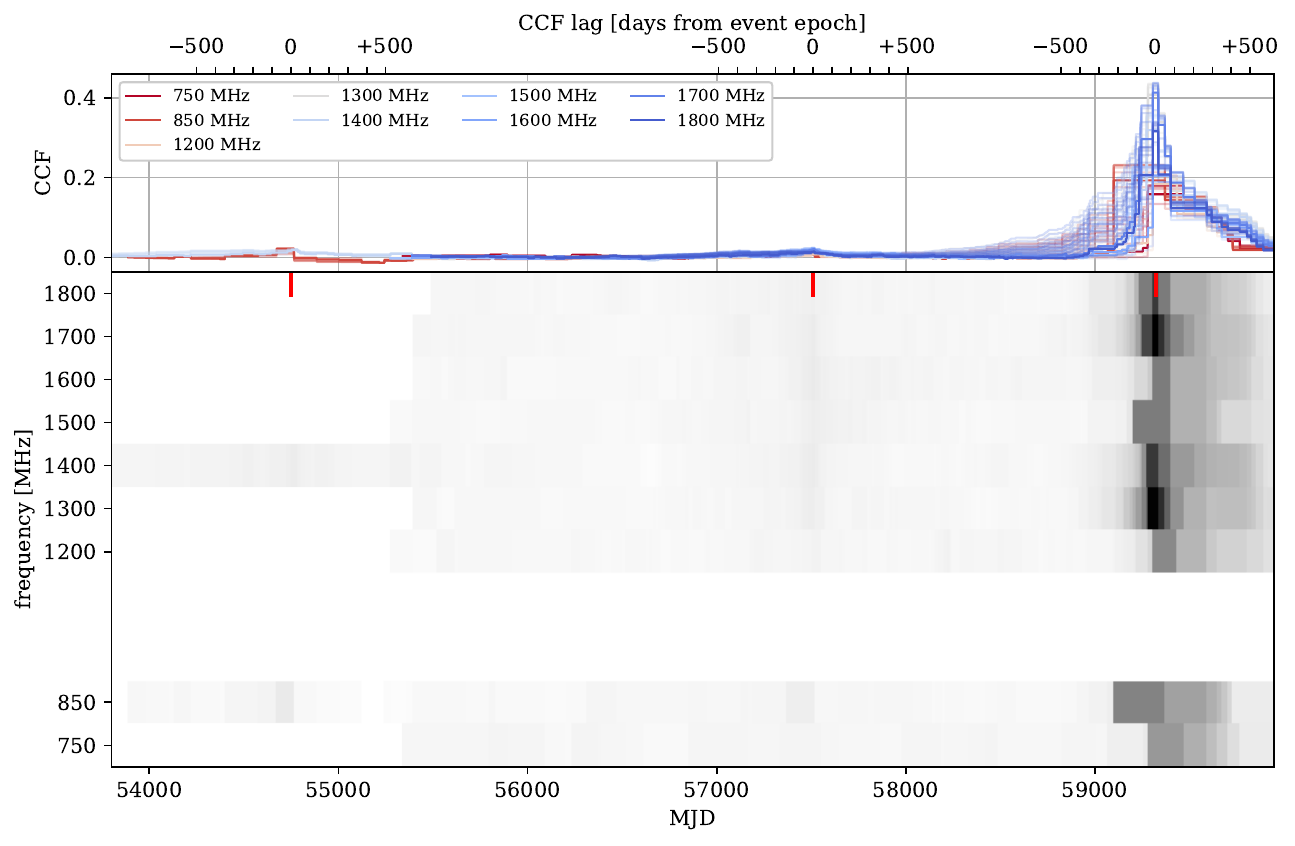}
    \caption{Top panel: CCF responses of PC 2 projection magnitudes of PSR J1713+0747 data with FRED templates with $\tau$ ranging from 50 to 300\,d in increments of 10\,d. Bottom panel: A heatmap of CCF responses over frequency with $\tau_0$ fixed at 100\,d. The epochs of known events are given by the red ticks and define the zero points along the top $x$-axis.}
    \label{fig:matchfilterheatmap}
\end{figure*}

\begin{figure*}
    \centering
    \includegraphics[width=\textwidth]{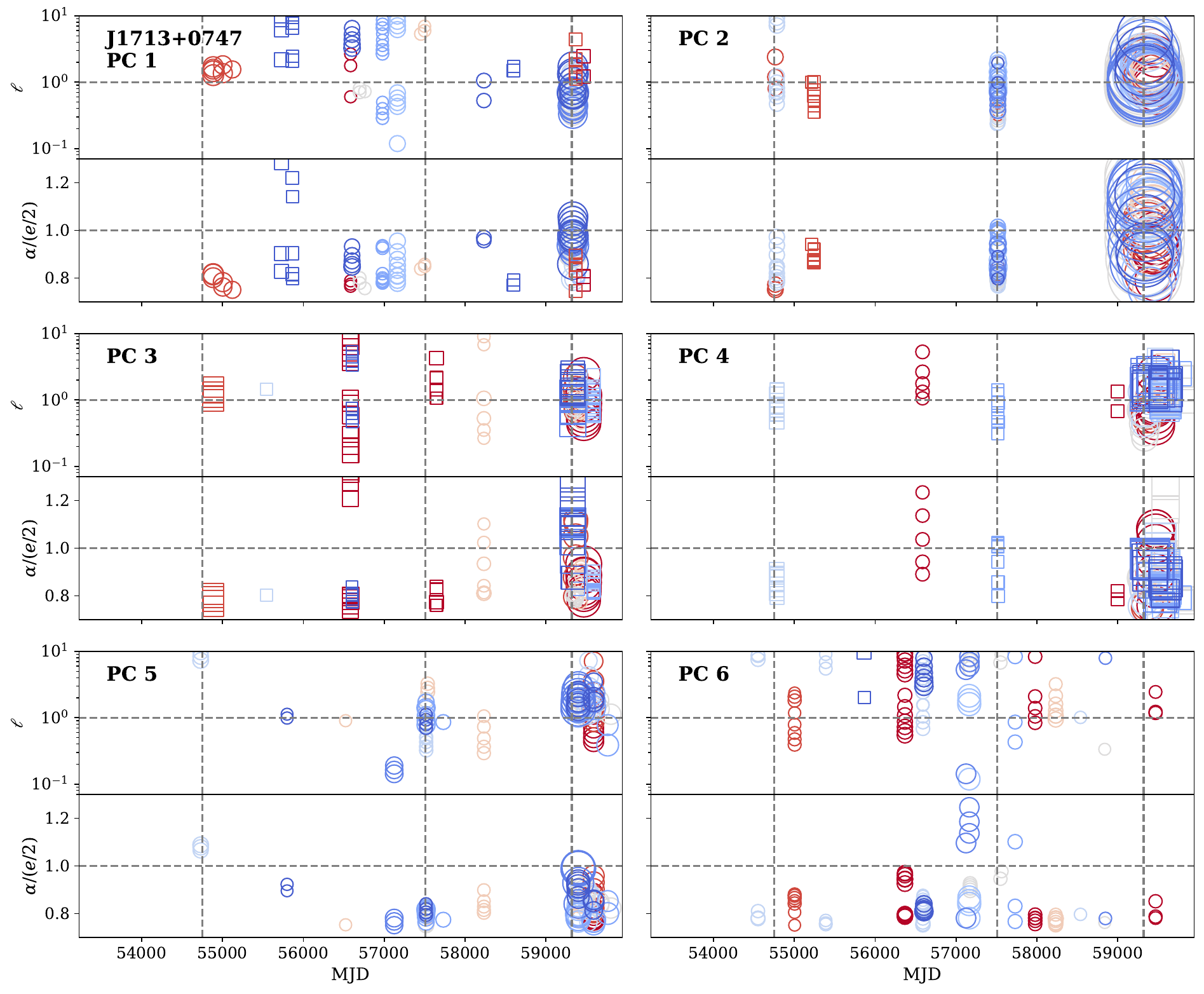}
    \caption{Time series of $\ell$ and $\alpha$ values calculated for pairs of maxima or minima between the FRED CCF and the DERF CCF for each of the first six PCs for each pulsar in our sample. Circles are pairs of maxima; squares are pairs of minima. The size of each marker gives the S/N of the FRED CCF extremum for that pair, though these S/N values are not normalized by eigenvalues on a per-PC basis as the panels in Figure \ref{fig:pca} are. Color denotes frequency in the same way as in Figure \ref{fig:pca}.
    The complete figure set (11 images) is available in the online journal.}
    \label{fig:6pcs_params}
\end{figure*}

We seek pulse shape change events of the type seen in 2021 April for PSR J1713+0747: a sudden pulse shape change followed by a gradual relaxation back to the original pulse shape. Since the decay in TOA residuals of the 2021 event is roughly exponential, we use a fast rise, exponential decay (FRED) template to filter for similar events in the $D_i$ time series. For each PC, and for each 100\,MHz subband, we cross-correlate the $D_i$ time series with positive FRED templates with a range of decay times $\tau_j$. This returns a cross-correlation function (CCF), where peaks correspond to good matches between the test shape and any similar structures in the data. We normalize the CCF response by the norm of the template vector. 

We assume an exponential decay rate for our analysis because the TOA residuals and DM residuals appear to follow FRED curves for the 2021 PSR J1713+0747 event, but other decay rates are possible. We test power law fits for the three PSR\,J1713+0747 events and reject them in favor of FRED curve fits on the basis of their residual rms. Searches seeking greater sensitivity to events with nonexponential decay rates may replace the FRED template in our analysis with another template.

As a screen against outliers, we run the same cross-correlation and extrema-finding algorithm with a time-reversed copy of our template: an exponential rise, fast decay shape (DERF; ``FRED'' backward). Single-epoch outliers can produce strong responses in the FRED CCF, but will produce statistically similar responses in the DERF CCF. On the other hand, true FRED events will have quantifiably different FRED and DERF responses. We demonstrate the difference with a toy model in Figure \ref{fig:matchfiltertoy}, where the top panel shows FRED and DERF responses to a FRED event in data and the bottom panel shows responses to a single-epoch outlier (a delta function). The simulated data are uniformly sampled, an approximation that we address in greater depth in Appendix \ref{app:nonuniform}. The top panel of Figure \ref{fig:matchfilterex} shows the FRED and DERF responses for the 1400\,MHz $D_2$ time series for PSR\,J1713+0747. The middle and bottom panels show decaying exponential fits to the epochs of the known events in that pulsar, which we further discuss in Section \ref{sec:J1713+0747}.

We measure a candidate event's adherence to the FRED template with two quantities: the time lag between the extrema in the FRED and DERF CCFs and the ratio of their amplitudes. In the remainder of this section, we derive the values of these quantities for the case of an idealized FRED event (the top panel of Figure \ref{fig:matchfiltertoy}). In Section \ref{sec:ranking}, we use these quantities to define the $\zeta$ score, an event candidate ranking metric.

Let $h(t) = \Theta(t)e^{-t/\tau_0}$, where $\Theta(t)$ is the Heaviside function, be a FRED template with amplitude 1 and decay time $\tau_0$. The cross-correlation of $h(t)$ with itself (that is, the autocorrelation of $h(t)$) is
\begin{align}
    A^{(+)}(t') &= \int dt\,h(t)\,h(t+t')\\
    &= e^{-t'/\tau_0}\int dt\,\Theta(t)\Theta(t+t')\,e^{-2t/\tau_0}.
\end{align}

Here, $(+)$ represents the forward pass of the FRED template; later, $(-)$ will denote the CCF with the DERF template. $A^{(+)}(t')$ is nonzero only when $\Theta(t) = \Theta(t+t') = 1$. When $t'$ is positive, the product $P(t, t') = \Theta(t)\Theta(t+t')$ is constrained by $\Theta(t)$; we have $P(t,t') = 1$ for all $t>0$. In this case, 
\begin{align}
    A^{(+)}(t'>0) &= e^{-t'/\tau_0} \int_0^\infty dt\,e^{-2t/\tau_0}\\
    &= \frac{\tau_0}{2}e^{-t'/\tau_0}.
\end{align}
When $t'$ is negative, $P(t,t')$ is constrained by $\Theta(t+t')$; we have $P(t,t') = 1$ for all $t > -t'$. In this case,
\begin{align}
    A^{(+)}(t'<0) &= e^{-t'/\tau_0} \int_{-t'}^\infty dt\,e^{-2t/\tau_0}\\
    &= \frac{\tau_0}{2}e^{t'/\tau_0}.
\end{align}
Thus we have
\begin{equation}
    A^{(+)}(t') = \frac{\tau_0}{2}e^{-|t'|/\tau_0},
\end{equation}
and the peak amplitude expected for a cross-correlation of a FRED event with a FRED template of equal $\tau_0$ (the FRED autocorrelation) is $A^{(+)}(0) = \tau_0/2$. Our normalization for the discrete case, including the case shown in Figure \ref{fig:matchfiltertoy}, includes a factor of $\langle\delta t_\text{samp}\rangle/T$, where $\delta t_\text{samp}$ is the sampling rate, $T$ is the timespan of observations, and the angle brackets denote an average to account for nonuniform sampling, further addressed in Appendix \ref{app:nonuniform}.

Now we treat the backward (DERF) pass. The cross-correlation of a FRED event with a DERF template is 
\begin{align}
    A^{(-)}(t') &= \int dt\,h(t)\,h(-t+t')\\
    &= e^{-t'/\tau_0}\int dt\,\Theta(t)\Theta(-t+t').
\end{align}
The product $\Theta(t)\Theta(-t+t')$ is nonzero only when $t>0$ and $t<t'$. Thus,
\begin{align}
    A^{(-)}(t') &= \Theta(t')\,e^{-t'/\tau_0}\int_0^{t'} dt\\
    &= \Theta(t')\,t'\,e^{-t'/\tau_0}.
\end{align}
The peak response of the DERF CCF occurs at $t'=\tau_0$. The amplitude at the peak is $A^{(-)}(\tau_0) = \tau_0/e$.

The lag time between the extrema of the FRED CCF and the DERF CCF is $\tau_0$ and the ratio of their amplitudes is $\alpha = A^{(+)}(0)/A^{(-)}(\tau_0) = e/2$. Since the lag time varies with $\tau_0$ and we test a range of decay times, we define the dimensionless lag as $\ell = \tau_\text{obs}/\tau_0$, where $\tau_\text{obs}$ is the observed lag between CCF responses for a template with decay time $\tau_0$. The expected response for a true FRED event is $\ell = 1$.

Our range of trial decay time values, $\tau_j$, is 50--400\,d in steps of 50\,d. We choose this range by inspection of the three known PSR\,J1713+0747 pulse shape change events; all three have decay times of order 100\,d. For PSR J1713+0747, we show a set of CCFs for different subaveraged frequency channels and FRED templates with different $\tau_j$ values in the top panel of Figure \ref{fig:matchfilterheatmap}, and a heatmap over frequency with fixed $\tau = 100$\,d in the bottom panel. 

We are not completely insensitive to events with decay times outside this range, though the observation schedule complicates searching for them. A FRED template with $\tau \ll 50$\,d will be approximately a delta function compared to the cadence of NANOGrav observations for most PTA pulsars, not easily distinguishable from the single-day outlier shape changes we typically attribute to instrumental or calibration errors. On the other hand, a template with $\tau \gg 400$\,d approaches the overall observation timescale, diminishing the significance of the pulse shape's stable baseline and biasing the PC calculation. Within our range, we find that the amplitude of the CCF response is not very sensitive on grid scales finer than $\sim$50\,d, resulting in very similar detection parameters $\ell$ and $\alpha$ over small changes in decay time. While we tested a $\tau_j$ range in increments of 10\,d, we conduct our full search over a 50\,d grid for speed.

PCA guarantees an orthogonal basis of eigenvectors, but an orthogonal basis with any vector replaced with its negative is still an orthogonal basis. A pulse shape change event may therefore manifest as a positive or a negative jump. We see this in the $D_2$ time series of PSR\,J1713+0747 and PSR\,J1643$-$1224: the events in PSR\,J1713+0747 are positive jumps while the one in PSR\,J1643$-$1224 is a negative jump. Practically speaking, the main effect of this is to require that we search not only for maxima but also for minima in our CCF time series. We impose a prominence threshold of 800\,d (twice the maximum $\tau_j$ in our search range); that is, if there are multiple local CCF maxima within 800\,d of each other, only the one with the greatest magnitude is retained, and the same for local minima.

\subsection{Candidate Ranking}\label{sec:ranking}

\begin{center}
\begin{table*}
\begin{threeparttable}
    \caption{Top-Ranked Candidate Events.}
    \begin{tabular}{l l l c | c c c c c | l}
        \hline
        \hline
        Pulsar  & Date & MJD & $\zeta_\text{min}$ & $\zeta_1$ & $\zeta_2$ & $\zeta_3$ & $\zeta_4$ & $\zeta_5$ & Notes\\
        \hline
J1713+0747&2021 Apr 16&59320\tnote{a}&0.05& 0.05& 0.08& 0.13& 0.07& 0.10& Known event. \\
J1713+0747&2016 May 2&57510\tnote{b}&0.08& --- & 0.08& 3.00& --- & 0.44& Known event. \\
J0030+0451&2015 Sep 25&57290&0.11& 0.39& --- & 0.11& 0.49& --- & Nearby event in interpulse.\\
B1937+21&2015 Jan 18&57040&0.16& 0.85& 0.16& 2.72& --- & --- & Scattering variability. \\
B1937+21&2020 Dec 7&59190&0.17& 0.54& 0.27& 0.17& 0.34& 0.36& Scattering variability. \\
J1600$-$3053&2008 May 24&54610&0.20& 0.60& --- &0.20& --- & --- & Anomalous DMX.\\
J1643$-$1224&2015 Feb 21&57074\tnote{c}&0.24& 0.55& 0.24& 5.79& 0.88& --- & Known event. \\
J1713+0747&2008 Oct 11&54750\tnote{d}&0.25& 0.75& 0.25& 0.52& --- & --- & Known event. \\
        \hline
    \end{tabular}
    \begin{tablenotes}
        \footnotesize
        \item[a] Reported by \citet{Xu21}.
        \item[b] Reported by \citet{Lam18}.
        \item[c] Reported by \citet{Shannon16}.
        \item[d] Reported by \citet{Demorest13}.
    \end{tablenotes}
    \label{tab:candidates}
\end{threeparttable}
\end{table*}
\end{center}

\begin{figure}
    \centering
    \includegraphics[width=\columnwidth]{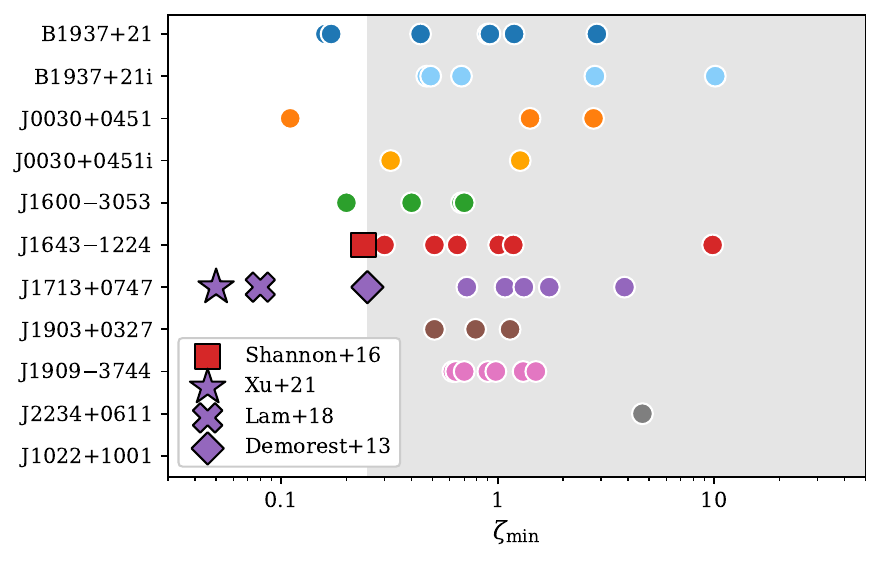}
    \caption{Distribution of $\zeta_\text{min}$ for all candidate events in each pulsar in our sample (see Tables \ref{tab:candidates} and \ref{tab:allcandidates}). The four known events are indicated with unique markers. Lower $\zeta_\text{min}$ values indicate better candidates, with events to the left of the gray shaded region considered significant.}
    \label{fig:zetadist}
\end{figure}

For an extremum to be considered as potentially belonging to a pulse shape change event, we require that it exceed an S/N threshold of 8. This threshold is calculated in the same way as for the $D_i$ time series (that is, using as a noise estimate the contiguous 1000\,d interval that minimizes the standard deviation), but it is calculated with respect to the CCF, so it is not redundant with the earlier S/N cut used to determine which profiles contribute to the PC calculation and $D_i$ time series.
Any extrema below this threshold are discarded at this stage of the analysis, simplifying the calculation of parameters for outlier rejection.

For each surviving FRED-DERF pair of extrema, we calculate the dimensionless lag $\ell$ and the amplitude ratio $\alpha$. As we show in the previous section, $\ell = 1$ and $\alpha = e/2$ for an ideal FRED event matched with FRED and DERF templates of the correct width. As a preliminary filter, we reject any extrema pairs for which the calculated quantities are too far from these fiducial values: any $\ell$ more than an order of magnitude from 1 and any $\alpha/(e/2)$ more than 0.25 from 1. For an idealized delta function, $\alpha/(e/2) \approx 0.736$, so this range excludes many of the single-epoch outliers in our data.

Any time-symmetric pulse shape change event will likewise have $\alpha/(e/2) \approx 0.736$, but small asymmetries may elevate $\alpha$ into our search range. In this case, $\ell$ is determined by the width of the event and in some cases may approximate 1, the value expected for a FRED-like event. An $\alpha/(e/2) \approx 0.736$ will still keep the $\zeta$ score out of the cutoff range used for Table \ref{tab:candidates} ($\zeta \approx e/2 - 1 \approx 0.359$ even if $n_{\rm chan} = N_{\rm chan}$), but these events may yet be astrophysical. We focus on FRED-like events in this work, but our method could be modified to use a Gaussian (or other time-symmetric) template instead of a FRED template for greater sensitivity to such events.

Figure \ref{fig:6pcs_params} shows $\ell$ and $\alpha$ for the time series of dot products for the first six PCs of each pulsar in our sample. Extrema pairs are only plotted here if they fall within our fiducial boundaries for $\ell$ and $\alpha$ and surpass the S/N threshold of 8. Many markers are plotted for a single epoch in most cases; this is typically because a single event candidate may occur in multiple frequency channels and/or because multiple $\tau_j$ produce similar CCF responses. To absorb this multiplicity and assign each event candidate a single epoch, we cluster the extrema pairs in time (MJD) using the \verb+scikit-learn+ \citep{scikit-learn} implementation of the mean shift clustering algorithm \citep{MeanShift} in 1D. We use a clustering bandwidth of 200\,d, roughly the midpoint of our $\tau_j$ search range. 

From each cluster, we choose the $\ell$--$\alpha$ pair that minimizes the distance from the fiducial values of 1 and $e/2$; we consider this pair to be the best detection over $\tau_j$ and frequency of the candidate event. We use this pair to calculate a candidate ranking score:
\begin{equation}\label{eq:zeta}
    \zeta = \frac{\sqrt{(\ell - 1)^2 + (\alpha - e/2)^2}}{n_\text{chan} / N_\text{chan}},
\end{equation}
where $n_\text{chan}$ is the number of frequency channels in which the event is detected and $N_\text{chan}$ is the number of frequency channels overall.\footnote{For most of our pulsars, $N_\text{chan} = 9$ for GUPPI and VEGAS data and 7 for PUPPI data due to the reduced sensitivity range. $N_\text{chan} = 2$ for ASP and GASP data due to the narrower bandpass. For PSR\,J1903+0327, $N_\text{chan}=6$ for PUPPI data and 1 for ASP data due to a lack of 430\,MHz observations for this pulsar. We consider candidates detected less than 200\,d, roughly the midpoint of our decay time search range, prior to the adoption of PUPPI or GUPPI to use the later receiver's $N_\text{chan}$ value.} Extrema pairs with $\zeta$ closer to 0 are considered better pulse shape change event candidates.

Where a single event is detected in multiple PCs, we keep the lowest $\zeta$ score, which we term $\zeta_\text{min}$. We report $\zeta_\text{min}$ and $\zeta$ scores for the first five PCs for our top-ranked event candidates in Table \ref{tab:candidates}. In Figure \ref{fig:zetadist}, we show the $\zeta_\text{min}$ distribution for each pulsar in our sample. A shaded region indicates the threshold for inclusion in Table \ref{tab:candidates}; the cutoff point is the 2008 \citep{Demorest13} event in PSR\,J1713+0747. A table of $\zeta$ scores for all event candidates is given in Appendix \ref{app:list}.

Of the eight candidates in Table \ref{tab:candidates}, four are those already known from \citet{Demorest13}, \citet{Shannon16}, \citet{Lam18}, and \citet{Xu21}. In Figure \ref{fig:knowncandidates}, we show snippets of the $D_i$ time series (i.e., from Figure \ref{fig:pca}) surrounding the epochs where these events were recovered. The epochs themselves are indicated by vertical dashed lines. The other four candidates do not correspond to previously reported pulse shape change events and are discussed further in Sections \ref{sec:B1937+21}, \ref{sec:J1600-3053}, and \ref{sec:J0030+0451}. In Figure \ref{fig:othercandidates}, we show snippets of the time series in which these events were detected with the best $\zeta$ scores. We also show the FRED templates fit to each detection.

\section{Results}\label{sec:results}

For each pulsar, the time series of normalized pulse profile residual dot products with the dataset's first five PCs, $\sigma_i D_i$, are shown in a figure in the Figure \ref{fig:pca} set (available in full in the online journal). As discussed in Section \ref{sec:pca}, some of our pulsars are brighter than others, so we choose an S/N cutoff for each pulsar according to the S/N distribution for its profiles. The same cutoff determines which profiles are included in the time series of residual dot products (Figure \ref{fig:pca}). 
The cutoff value for each pulsar is given in Table \ref{tab:snrcuts}; although interpulses are typically much lower in S/N than main pulses, we use a single cutoff for each pulsar whether it has an interpulse or not. In this section, we discuss the salient features of each time series.

\subsection{PSR J1713+0747}\label{sec:J1713+0747}

\begin{figure*}
    \centering
    \includegraphics[width=0.48\textwidth]{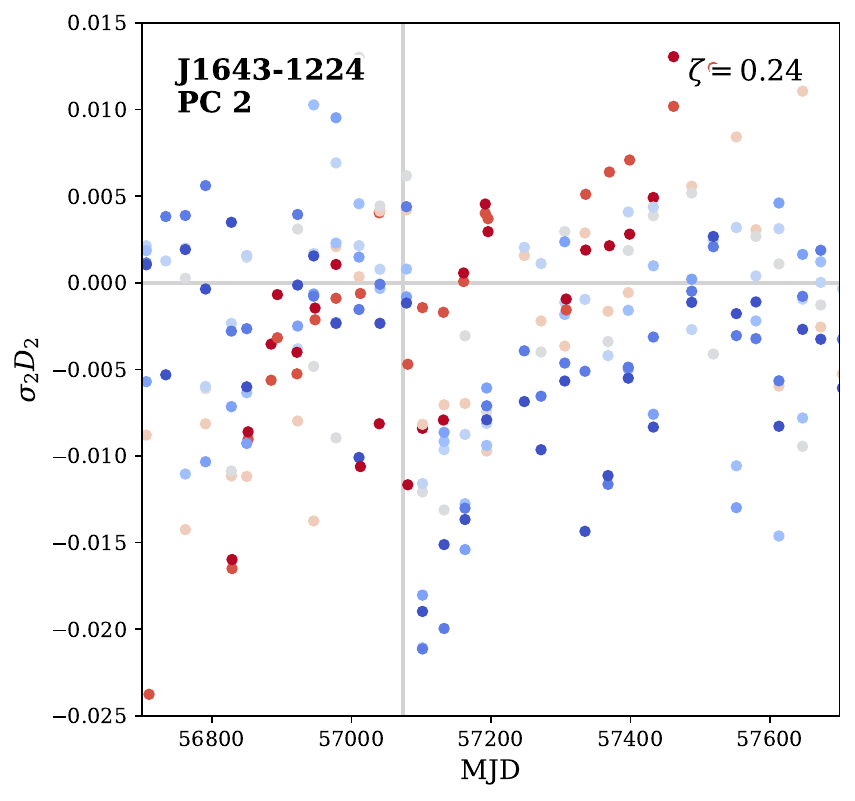}
    \includegraphics[width=0.48\textwidth]{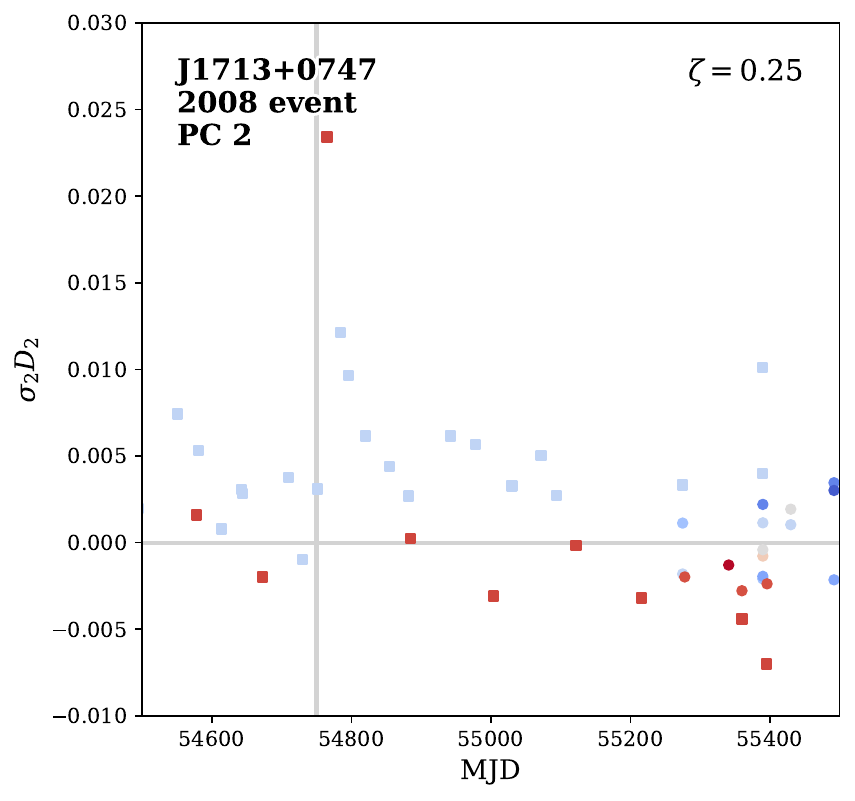}
    \includegraphics[width=0.48\textwidth]{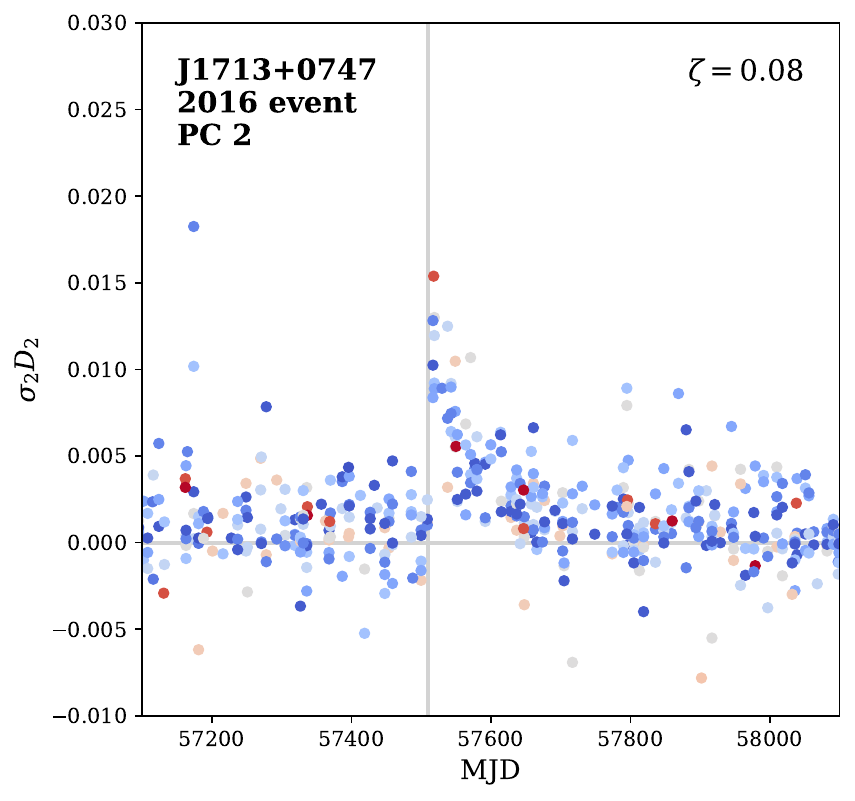}
    \includegraphics[width=0.48\textwidth]{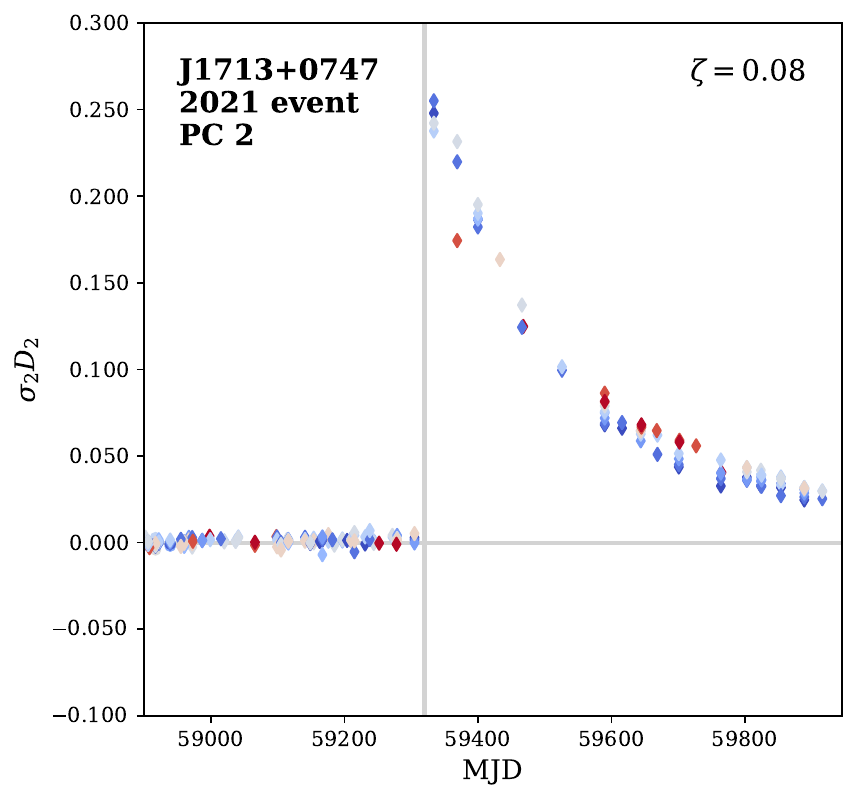}
    \caption{
    The four known pulse shape change events as seen in our PCA time series. Note the factor of 10 difference in $y$-axis scale for the bottom-right panel, showing the 2021 event in PSR\,J1713+0747, versus all other panels; despite this factor of 10, the $\zeta$ scores for the 2016 and 2021 events in PSR\,J1713+0747 are almost the same because $\zeta$ has no direct S/N dependence. In all panels, the scatter plots are snippets of $D_i$ time series in the Figure \ref{fig:pca} figure set, with color coding frequency and marker coding back end in the same way as in that figure set.}
    \label{fig:knowncandidates}
\end{figure*}

\begin{figure*}
    \centering
    \includegraphics[width=0.48\textwidth]{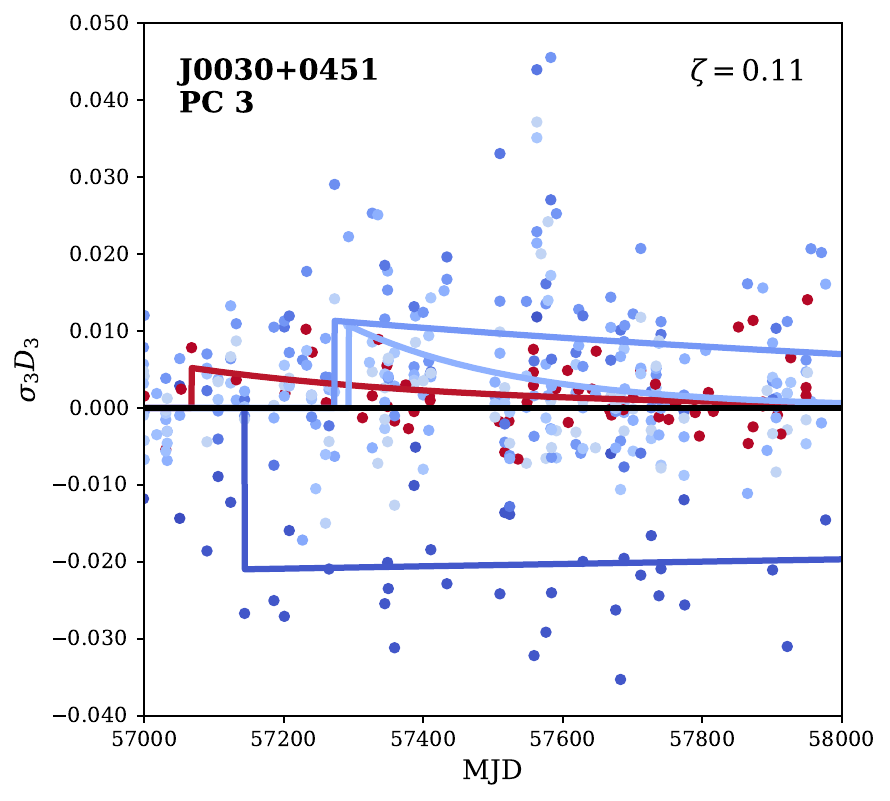}
    \includegraphics[width=0.48\textwidth]{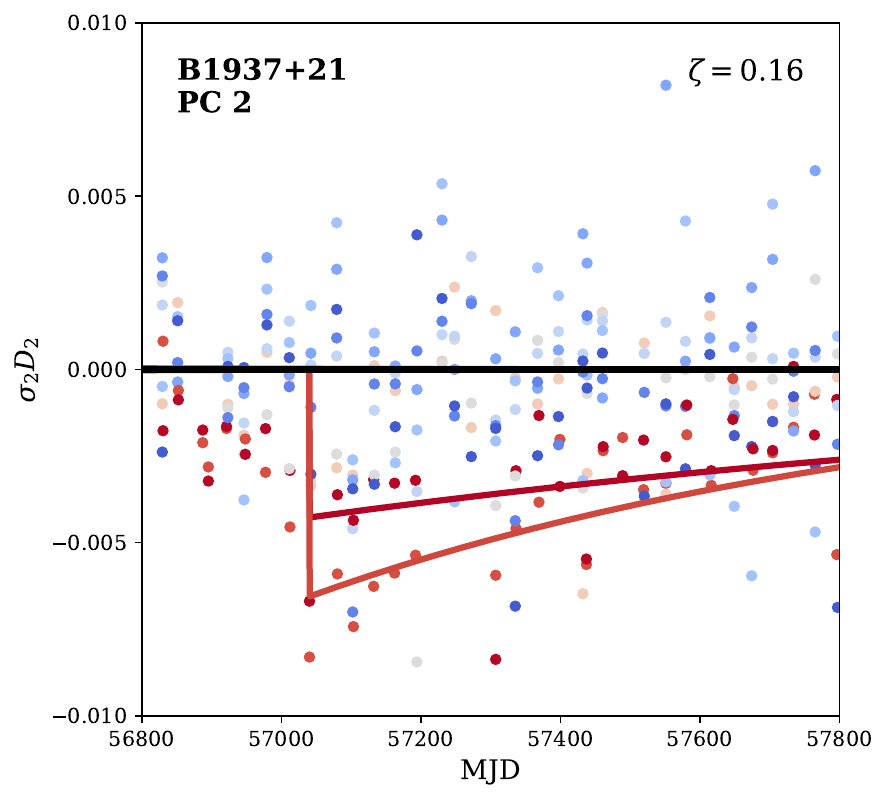}
    \includegraphics[width=0.48\textwidth]{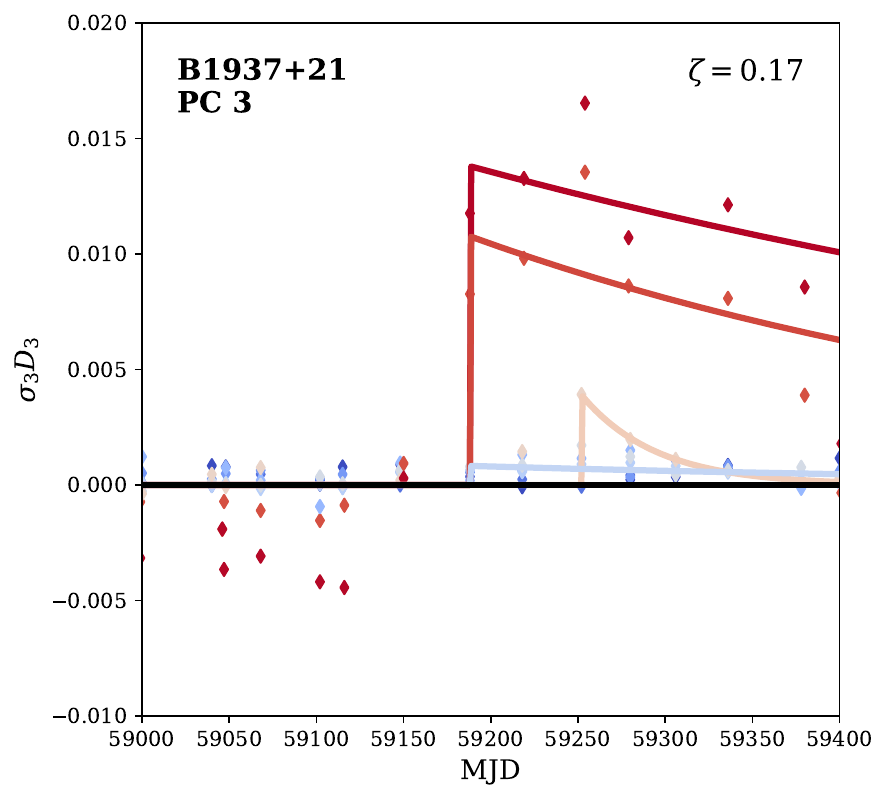}
    \includegraphics[width=0.48\textwidth]{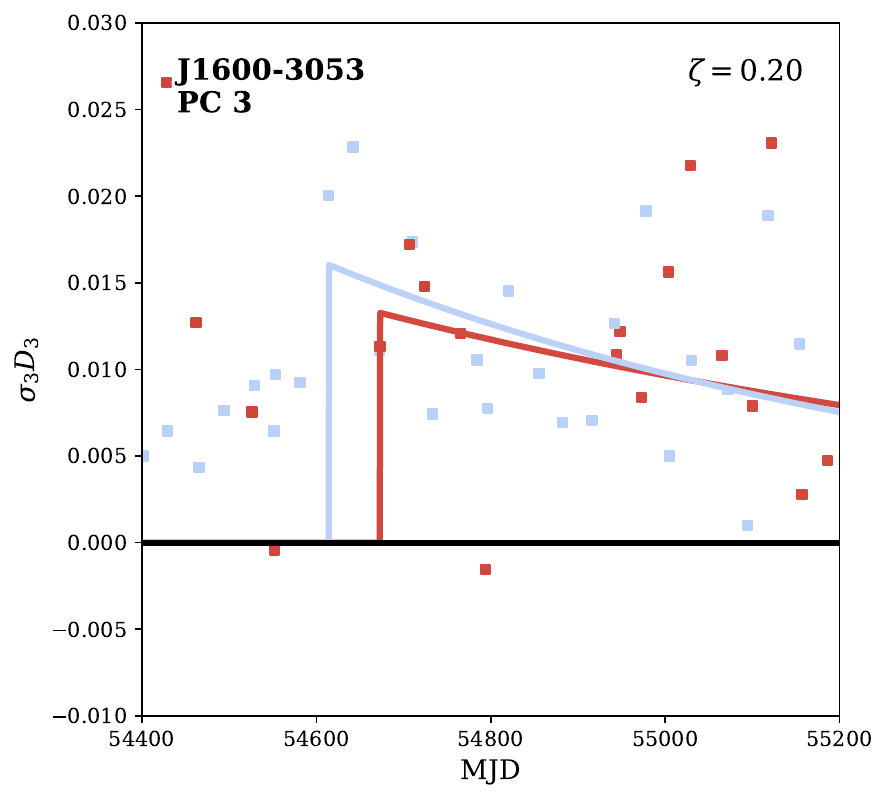}
    \caption{The four pulse shape change event candidates found by our automated search with $\zeta < 0.25$, the threshold set by the 2008 event in PSR\,J1713+0747. FRED shapes are overlaid in each case with the best-fit amplitude and decay time for each 100\,MHz frequency channel in which the candidate was detected. Like in Figure \ref{fig:knowncandidates}, in all four panels of this figure, the scatter plots are snippets of $D_i$ time series in the Figure \ref{fig:pca} figure set, with color coding frequency and marker coding back end in the same way as in that figure set.}
    \label{fig:othercandidates}
\end{figure*}

\begin{figure}
    \centering
    \includegraphics[width=\columnwidth]{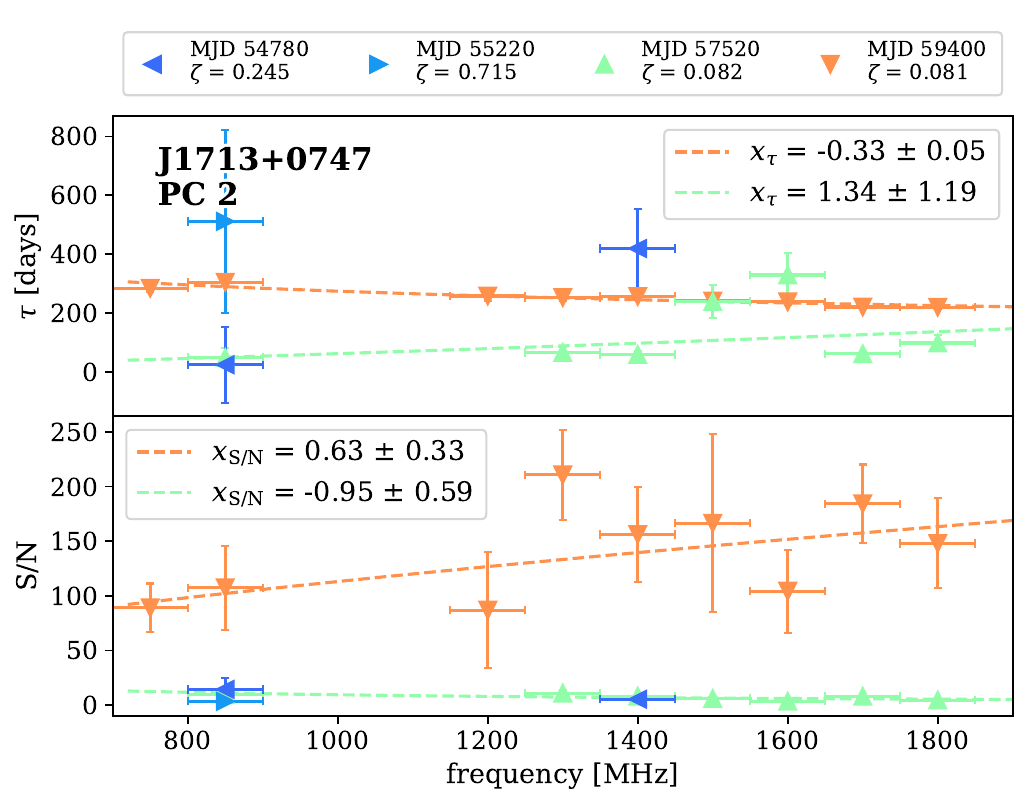}
    \caption{Spectral dependence of $\tau$ and S/N for each event isolated by the 1D mean-shift clustering algorithm in PC 2 of PSR J1713+0747 data. 
    For greater $y$-axis resolution, the $\tau$ and S/N shown here are the result of FRED template fitting to the $D_i$ time series around the detection epoch. Vertical error bars indicate the $1\sigma$ error in this fit, while horizontal error bars indicate the subband width in frequency.
    Power laws are fit to the 2016 and 2021 events.}
    \label{fig:specdep}
\end{figure}

The 2021 April event reported by \citet{Xu21} is clearly visible in all five panels of Figure \ref{fig:pca}.1. In the second panel, the timing events reported by \citet{Demorest13} and \citet{Lam18} are likewise visible, albeit at significantly lower amplitude. The variance captured by PC 2 appears to be dominated by the pulse shape change events (indeed, this is the panel where the events occur at greatest S/N), but contains very little chromatic variation. This supports the result of \citet{Lin21}: all three timing events correspond not only to DM variations, explainable through changes in the ISM along the line of sight, but to intrinsic pulse shape changes. 

Our method returns four candidate events in PC 2: the three known pulse shape change events, which are detected at MJDs 54780, 57520, and 59400, and a fourth candidate at MJD 55220, which we take to be spurious on the basis of its high $\zeta$ score.

We fit FRED curves to the 3\,yr of PC 2 dot products following each event and report their parameters, decay time $\tau$ and S/N, as a function of frequency in Figure \ref{fig:specdep}. The baseline used for each S/N calculation is the contiguous 1000\,d interval that minimizes the rms noise of the time series at that frequency. The spurious event is detected in only one 100\,MHz frequency channel, contributing to its high $\zeta$ score. The 2008 event is detected in both of the channels available at the time of its observation, when the GASP back end was in use. The 2016 and 2021 events are detected in sufficiently many channels to fit a power law to their $\tau$ and S/N values. In both cases for each event, the power law is fairly flat, supporting the observation that PC 2 captures mostly achromatic pulse shape variation. However, the lowest S/N occurring at 800\,MHz runs counter to the finding of \citet{Jennings24} that the event affected the pulsar's timing most drastically around this frequency. We emphasize that Figure \ref{fig:specdep} gives the FRED fit parameters only for the events' projections along the second PC; other PCs show more chromatic variation in this event, which may account for the missing power at 800\,MHz.

The epochs at which we find the three real events differ from those reported by \citet{Demorest13}, \citet{Lam18}, and \citet{Xu21} by less than a few times the observing cadence; we use our method's detected epochs in Figure \ref{fig:specdep} but the authors' reported epochs in Table \ref{tab:candidates}.

Figure \ref{fig:matchfilterex} shows the FRED fits for the three known events overplotted on the 1400\,MHz $D_2$ time series. The simplest method for mitigating these events in a timing analysis may be to excise the observations containing those events from the dataset. A criterion for excision is that the event amplitude at the trailing edge of the excised time block should be no more than $\epsilon\sigma_{\rm n}$, where $\sigma_{\rm n}$ is the rms noise in the baseline of the time series $D_i$. (The time series to choose or the method for averaging over several is open to interpretation; as a heuristic, we choose the 1400\,MHz $D_2$ time series shown in Figure \ref{fig:matchfilterex}.) The length of the excised time block is 
\begin{equation}
    T_{\rm exc} = \tau \ln(A/\epsilon\sigma_{\rm n}),
\end{equation}
where $A$ is the amplitude of the event. Using the FRED fits in Figure \ref{fig:matchfilterex} and choosing $\epsilon = 1$, we find $T_{\rm exc}$ values for the three events of $\sim$760\,d, 130\,d, and 1310\,d, respectively. (The exclusion time for the third event extends beyond the end of our dataset.) The total exclusion time is $\sim$2200\,d ($\sim$6\,yr).

\subsection{PSR J1643$-$1224}\label{sec:J1643-1224}

PSR\,J1643$-$1224 was shown by \citet{Shannon16} to have undergone a pulse shape change event around MJD 57074 (2015 February 21) due to a disturbance of the pulsar's magnetosphere. As discussed in Section \ref{sec:intro}, this event was strongest at 3\,GHz, weaker at 1.5\,GHz, and undetectable at 600\,MHz when observed with the 64\,m telescope Murriyang at Parkes Observatory. NANOGrav does not observe PSR\,J1643$-$1224 above 2\,GHz, but we expect to be able to detect the 2015 event in data from the GBT's $L$-band receiver.

We recover the 2015 event in the 1400, 1700, and 1800\,MHz frequency channels, though in our analysis it is absent from the lower end of the $L$-band data as well as at 800\,MHz. This has a minor effect on the event's $\zeta$ score, since its spectral occupancy is only three of the available nine frequency channels in this dataset.

\citet{Brook18} show long-term pulse profile variation in this pulsar at 800\,MHz through their Metric F, a measure of systematic variability as identified by a Gaussian process regression model compared to noisy variability. PSR\,J1643$-$1224 is known to have an H\,\textsc{ii} region along its line of sight \citep{Ocker24}, which contributes to its DM exceeding the expected DM for its distance. \citet{Brook18} conclude that refractive interstellar scintillation (that is, scattering variability) may be responsible for the long-term pulse shape variation, an explanation consistent with the presence of the H\,\textsc{ii} region. However, as we address further in Section \ref{sec:B1937+21}, there is no known H\,\textsc{ii} region along the line of sight to PSR\,B1937+21, the other pulsar in which \citet{Brook18} identify strong long-term scattering variability, suggesting that variation of this type can arise from other causes. We explore instrumental and solar systematics as potential causes in Section \ref{sec:B1937+21}, but favor an extrasolar origin. The chromaticity of the long-term variability strongly suggests an ISM effect along the line of sight and not an effect intrinsic to the pulsar's emission mechanism, but we cannot rule out the latter completely. Still, for brevity, we will refer to this behavior as ``scattering variability'' in the rest of this paper.

\citet{Maitia03} report an extreme scattering event in PSR\,J1643$-$1224 data at 1.28 and 1.41\,GHz occurring over 3\,yr ($\sim$1996--1999), which they interpret as the passage of a fully ionized cloud across the line of sight. This event occurred before NANOGrav observations commenced, so we do not recover it in our analysis, nor do we find any similar irregularities above 1\,GHz in NANOGrav's 15\,yr dataset.

We do recover both the magnetospheric pulse shape change event reported by \citet{Shannon16} at MJD 57074 and the scattering variability seen by \citet{Brook18}. Our recovery of the scattering variability is based mainly on a visual inspection of the time series at 800\,MHz; our matched-filtering method does not detect it at low $\zeta$. We recover no FRED-type events other than the 2015 event.

It is beyond the scope of this work to implement methods for mitigating the effects of pulse shape change events on the NANOGrav timing pipeline, but we point out that wider frequency coverage may permit future timing to avoid the worst of an event if it is isolated to a fraction of the band. For example, narrowband timing of PSR\,J1643$-$1224 would appear to avoid both the 2015 event and the slow variation if it is limited to frequencies between 1100 and 1400\,MHz. The challenge is that as yet, from our sample of four MSP pulse shape change events, we cannot predict prior to observing an event the range of frequencies over which it will occur, making it impossible to plan observations to avoid them. For now, the best course seems to be to observe these pulsars over as wide a frequency range as possible.
The PSR\,J1643$-$1224 event affects frequencies between 1.4 and 3\,GHz (or higher) and the 2021 event in PSR\,J1713+0747 affects frequencies between 0.6 and 1.9\,GHz \citep[at least; see][]{Jennings24}. Observations of most NANOGrav pulsars currently occur over a narrower range than this. However, observations up to 4\,GHz with the VLA and the new ultra-wideband receiver (UWBR) on the GBT \citep{UWBR20, UWBR23} will significantly extend the typical range for the 20\,yr and other future NANOGrav datasets. A possible outcome is the mitigation for timing of future pulse shape change events.

\subsection{PSR J1909$-$3744}\label{sec:J1909-3744}

\begin{figure}
    \centering
    \includegraphics[width=\columnwidth]{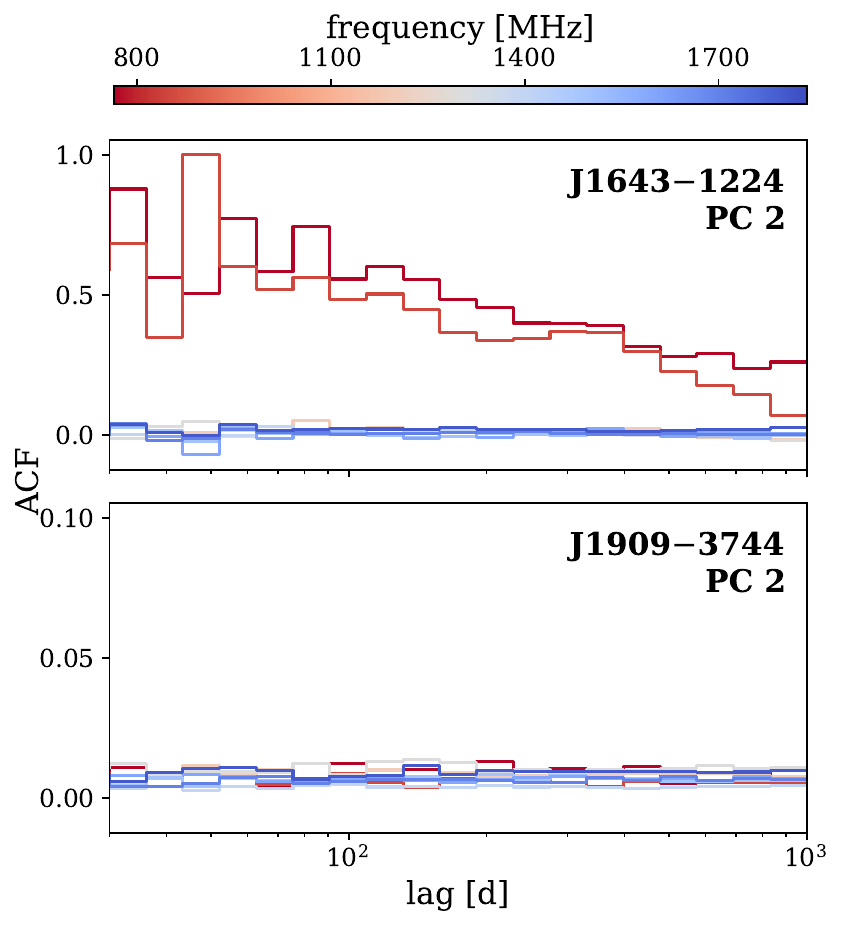}
    \caption{
    ACFs of PC 2 in (top panel) PSR\,J1643$-$1224 and (bottom panel) PSR\,J1909$-$3744. All ACFs in both panels are normalized to the maximum in the 850\,MHz PSR\,J1643$-$1224 ACF. PC 2 in PSR\,J1643$-$1224 shows long-term variability in the lower frequencies, reflected in the elevation of its ACF at low lag. Although PC 2 in PSR\,J1909$-$3744 has a visible positive skew, indicating non-Gaussianity, the ACF is flat for all frequencies, displaying no evidence for a red noise process. Note the factor of 10 difference between the $y$-axis scales.}
    \label{fig:acf}
\end{figure}

PSR\,J1909$-$3744 is the NANOGrav pulsar with the lowest weighted rms of post-fit TOA residuals \citep{NG15OneByOne}, an exemplar of ideal timing both for its timing stability and for its unusually simple pulse shape. 
We expect to find no pulse shape change events for this pulsar, and indeed we find none below $\zeta = 0.62$. 

Three of the seven PSR\,J1909$-$3744 candidate events show $\zeta < 1$ responses in PC 2, whose dot product time series has a positive-skewed distribution even to the eye. A look at the PCs (Figure \ref{fig:comps}.3) shows obvious structure in the first three of them, though the amplitude of the structure is lower for PC 2 than for the other two. The ``W'' shape in the residual is characteristic of pulse narrowing or an increase in amplitude.

The autocorrelation function (ACF) of the PSR\,J1909$-$3744 $D_2$ time series is shown alongside that of PSR\,J1643$-$1224 in Figure \ref{fig:acf}. For the latter pulsar, in which we observe strong scattering variability around 800\,MHz, the ACF is elevated at low lag at these frequencies. For PSR\,J1909$-$3744, the ACF is essentially flat at all frequencies, suggesting that despite the $D_2$ time series's positive skew, there is no underlying red noise process.

\subsection{PSR J2234+0611}\label{sec:J2234+0611}

PSR\,J2234+0611 is the second-least noisy pulsar (i.e., it has the second-lowest weighted rms of post-fit TOA residuals) in the NANOGrav 15\,yr PTA. It has been observed for considerably less time than the least noisy pulsar, PSR\,J1909$-$3744, which results in difficulties in PCA described further in Section \ref{sec:J1022+1001}. Still, we include it in our sample so as to have a second highly significant (and expected null) test case. Since we have far fewer epochs, we lower our S/N threshold for this pulsar to 15, lower than for any other in our sample but PSR\,J1903+0327, to ensure we have enough pulse profiles for PCA (see Table \ref{tab:snrcuts}). Like with PSR\,J1909$-$3744, we find no low-$\zeta$ pulse shape change event candidates using our automated search method, nor any anomalous time ranges by visual inspection.

\subsection{PSR B1937+21}\label{sec:B1937+21}

\begin{figure*}
    \centering
    \includegraphics[width=0.9\textwidth]{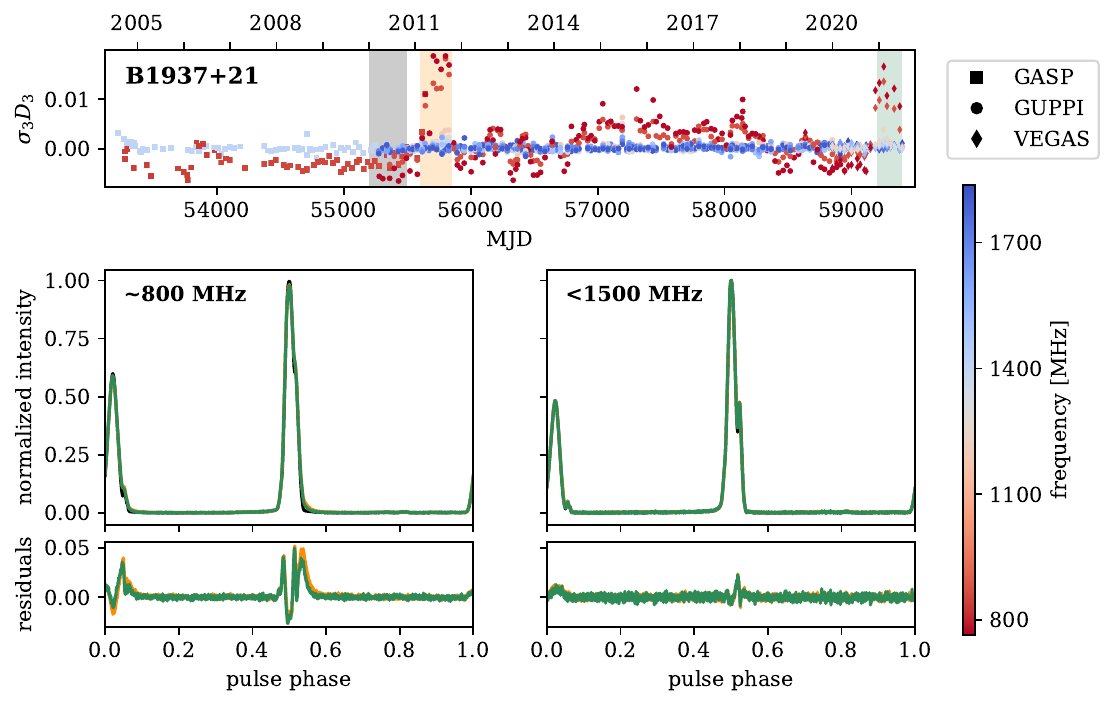}
    \caption{Top panel: Time series of projection magnitudes of PSR\,B1937+21 pulse profile residuals along their third PC. As in Figure \ref{fig:pca}.5, the dot products are normalized by $\sigma_3$, the square root of the corresponding eigenvalue. Regions of interest are indicated by the black, orange, and green vertical bars. Except for the bars, this plot is equivalent to the third row of Figure \ref{fig:pca}.5. Bottom left panel: Comparison of pulse profiles at $\sim$800\,MHz from epochs in the black (``baseline'') region of the top panel to those from the orange and green (``peak'') regions. Each curve is the average of all 800\,MHz pulse profiles in the corresponding region of the top panel; the three curves lie almost on top of each other. The residuals are also shown, demonstrating that the pulse shapes in the peak regions deviate from that in the baseline in almost exactly the same way. Bottom right panel: Identical to the bottom left panel, except that all $L$-band profiles up to 1500\,MHz from each region are averaged instead of all 800\,MHz profiles.}
    \label{fig:B1937+21:regions}
\end{figure*}

PSR\,B1937+21, the first MSP discovered \citep{Backer82}, is one of the brightest pulsars in the NANOGrav PTA. \citet{Brook18} report long-term variability in PSR\,B1937+21 at the highest level of any pulsar in their sample, which contained 38 of the 45 pulsars in the NANOGrav 11\,yr dataset; seven pulsars were excluded for having too few observations at the time.

We recover this long-term variability in a similar form to that of PSR\,J1643$-$1224, clearly visible in the dot products of pulse profile residuals with all of the first five PCs, shown in Figure \ref{fig:pca}.5. As with PSR\,J1643$-$1224, \citet{Brook18} interpret the variability as a scattering effect.

A notable feature in all panels of Figure \ref{fig:pca}.5 is a repeated shape change in the $\sim$800\,MHz pulse profiles: their shape varies slowly over the course of $\gtrsim$1\,yr between late 2010 and early 2012, and again nearly 10\,yr later, between mid-2020 and late 2021. Because the variation attains roughly the same amplitude in each $D_i$ time series, we expect the associated pulse profile shape changes to be very similar. The bottom-left panel of Figure \ref{fig:B1937+21:regions} shows the extent of the similarity: when comparing the average 800\,MHz pulse profile from each of the pulse shape change regions to the average profile in another region of the time series, the two pulse shape changes deviate almost identically despite occurring nearly 10\,yr apart. This effect also occurs to a lesser extent at higher frequencies, as shown in the bottom-right panel of Figure \ref{fig:B1937+21:regions}. Only profiles between 1200 and 1500\,MHz were averaged for the bottom-right panel; the effect becomes negligible at the higher end of the $L$-band receiver's range.

As both the bottom panels of Figure \ref{fig:B1937+21:regions} show, the same effect is seen in PSR\,B1937+21's interpulse. The first five $D_i$ time series for the interpulse are shown in Figure \ref{fig:pca}.6. 

We consider three possible sources of the 10\,yr repeated variation: instrumental errors, astrophysical effects local to the Solar System, and extrasolar effects, including interstellar scattering and variability intrinsic to the pulsar.
Below, we address these possibilities in greater detail.

\subsubsection{Instrumental Errors}

\begin{figure*}
    \centering
    \includegraphics[width=\textwidth]{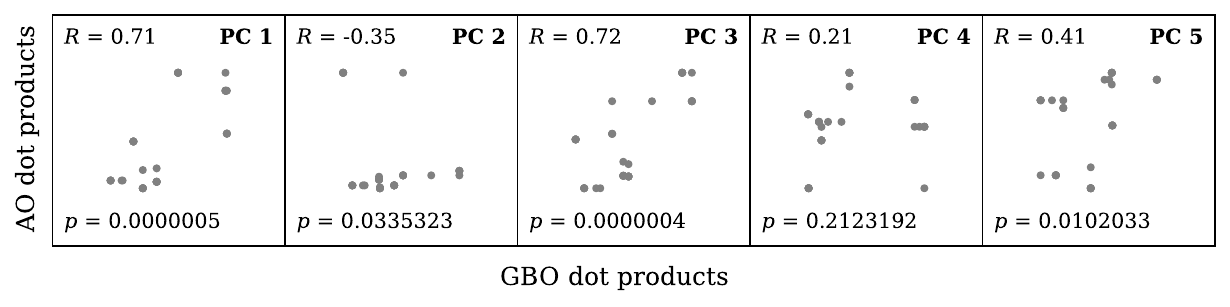}
    \caption{Scatter plots showing the correlations between PC magnitudes (dot products) for PSR\,B1937+21 pulse profiles recorded with the GBT and Arecibo at 1400\,MHz between MJDs 55600 and 55850 (the orange band of Figure \ref{fig:B1937+21:regions}). Only pulse profiles between 1350 and 1450\,MHz are shown due to the limitations of the ASP back end, which was in use at Arecibo at the time. Correlation coefficients ($R$) and $p$-values are given in each plot.
    }
    \label{fig:AOcorr}
\end{figure*}

\citet{Brook18} discuss known instances of instrumental error in GUPPI and PUPPI data for PSR B1937+21. Among the epochs typically excised from timing analysis are the MJD 57083--57263 range from PUPPI data, which coincided with elevated RFI activity before the installation of a new filter at Arecibo, and MJD 55977 from GUPPI data, one example of a day on which observations were calibrated using only a noise diode and not an on-sky flux calibrator source. As discussed in Section \ref{sec:obs}, we excise the latter from our analysis, too.

The standard NANOGrav data reduction pipeline assumes orthogonality of the linear feeds and no cross terms, but deviations from this assumption can affect pulse profile shapes and therefore, to some extent, timing \citep{Heiles01, vanStraten04}. Recalibration of pulsar timing data using the Mueller matrix for each detector has been implemented by \citet{Gentile18}, \citet{Wahl22}, and \citet{Dey24}, with the result that polarization profiles for many NANOGrav pulsars are publicly available. However, the analyses of \citeauthor{Gentile18} and \citeauthor{Dey24} relied in part on long-term observations of PSR B1937+21 as a standard of well-behaved polarization, making it difficult to apply their analyses to PSR\,B1937+21 itself. \citet{Wahl22} publish epoch-averaged polarization profiles for PSR\,B1937+21 derived from GBT observations of PSR\,B1929+10 with the 800\,MHz receiver and of PSR\,J1022+1001 with the $L$-band receiver.

Using the rotation measures of 820\,MHz pulse profiles, \citeauthor{Wahl22} found that the strength of the magnetic field along the line of sight to PSR\,B1937+21 varies sinusoidally with a period of $\sim$1\,yr. While many 1\,yr periodic trends in pulsar timing data are rightfully ascribed to the Sun, as noted in Section \ref{sec:B1937+21}, the line of sight to PSR\,B1937+21 never comes closer than $\sim$40$^\circ$ to the Sun on the sky. We do observe a pseudoperiodic ringing in some PCs for this pulsar at a lower level than the two major events, but as we discuss in Section \ref{sec:solar}, the period is $\sim$1.6\,yr, inconsistent with the 1.00$\pm$0.03\,yr magnetic field variation reported by \citet{Wahl22}. The eight pulsars in our sample with lines of sight that approach the Sun more closely (in some cases within a few degrees) do not display the same behavior, even PSR\,J1600$-$3053, PSR\,J1643$-$1224, and PSR\,J1713+0747, which were found by \citeauthor{Wahl22} to have sinusoidally varying line-of-sight magnetic field strengths with comparable or greater amplitude.

At 820\,MHz, \citeauthor{Wahl22} also report epoch-to-epoch variation of up to 18\% in the linear polarization of the PSR B1937+21 interpulse (i.e., after complete Mueller matrix calibration). In the $D_i$ time series in Figure \ref{fig:pca}, PCs are calculated from the main pulse and interpulse regions separately; the variation in the main pulse is shown in Figure \ref{fig:pca}.5 and the interpulse in Figure \ref{fig:pca}.6. 

\cite{Brook18} note that long-term pulse profile variability in PSR J1643$-$1224 has been observed using both the GBT and Murriyang, and accordingly they reject instrumental effects as a source of this variation. In the case of PSR B1937+21, NANOGrav observations were taken using both the GBT and the Arecibo Telescope, but only $L$-band and $S$-band observations were taken at Arecibo, i.e., none at the sub-GHz frequencies at which we see the strongest long-term pulse profile variation in GBT data. 

As we show in Figure \ref{fig:B1937+21:regions}, the pulse profile variation does also occur above 1\,GHz in GBT data, albeit at a much lower level. To check for correspondence between these data and those from Arecibo, we plot their PC dot products against each other using a window of 30\,days to match observation epochs. The dot products for the first five PCs are shown in Figure \ref{fig:AOcorr}. They exhibit moderate correlations in the first and third PCs, with coefficients of $\sim$0.7); the low $p$-values support the correlations' significance. The spread in the scatter plots is not unexpected given the low magnitude of the pulse shape variation above 1\,GHz.

On the basis of these correlations, combined with the fact that comparable variation does not seem to occur at 820\,MHz for any of the other NANOGrav pulsars observed with the GBT over the same period, we rule out instrumental effects as a source of the variation.

\subsubsection{Solar Systematics}\label{sec:solar}

\begin{figure}
    \centering
    \includegraphics[width=\columnwidth]{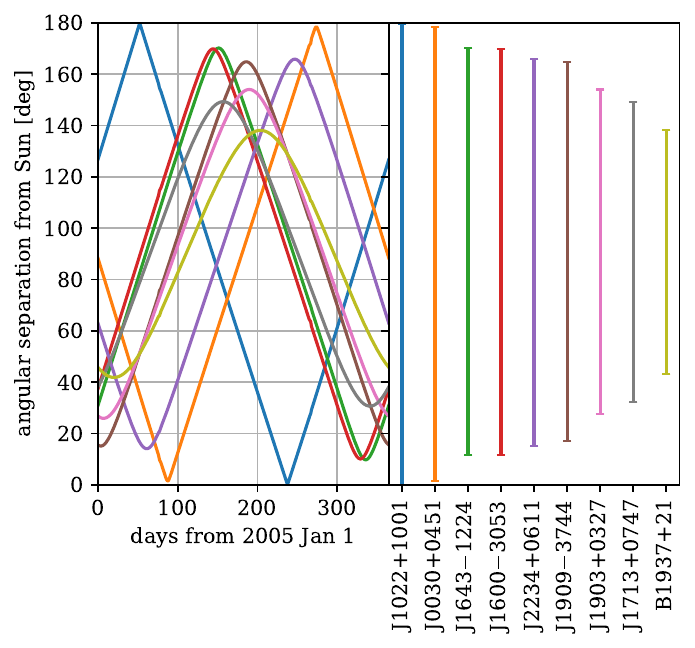}
    \caption{\emph{Left:} Angular separation of each pulsar in our sample from the Sun over 1\,yr. \emph{Right:} Range of angular separations for each pulsar. The coordinates used for each pulsar are those given in their names (e.g., $\alpha = 17^\mathrm{h}13^\mathrm{m}00^\mathrm{s}$ and $\delta = 07^\circ47'00''$ for PSR J1713+0747) and are inaccurate by up to a few degrees (for PSR B1937+21). The solar ephemeris is due to JPL Horizons.}
    \label{fig:solarelongation}
\end{figure}

\begin{figure}
    \centering
    \includegraphics[width=\columnwidth]{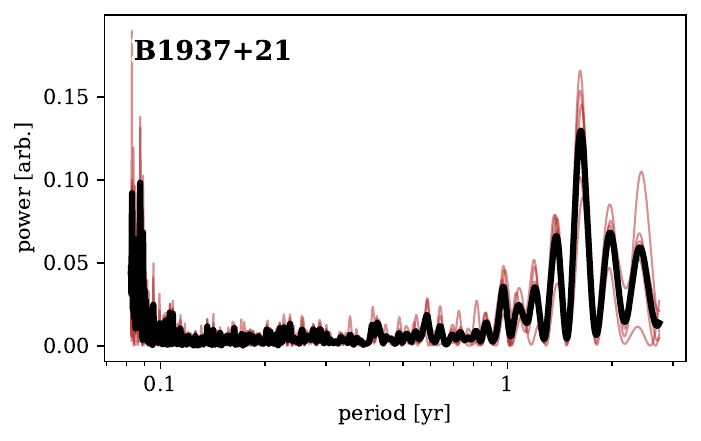}
    \caption{
    Lomb--Scargle periodograms of the 800\,MHz dot products for each of the first five PCs of PSR\,B1937+21 (in red) and the average of the five (in black). The peak on the right-hand side is at $\sim$1.6\,yr.}
    \label{fig:lombscargle}
\end{figure}

Pulsar timing is known to be affected by solar activity when the line of sight to the pulsar is close ($\lesssim$ a few degrees) to the Sun \citep{You12}. Some pulsars exhibit yearly DM increases, corresponding to the period when the solar wind in the Sun's immediate vicinity causes the most extreme overdensity of free electrons along the line of sight. The $\sim$11\,yr solar cycle is a tempting scapegoat for the $\sim$10\,yr periodicity of this pulse shape variation (if periodic it is, and not simply a two-time occurrence), but as shown in Figure \ref{fig:solarelongation}, PSR B1937+21 never gets closer than $\sim$40$^\circ$ to the Sun on the sky. In itself this does not guarantee no solar influence---PSR J1713+0747, for example, has obvious yearly DM increases despite its line of sight never approaching the Sun closer than $\sim$30$^\circ$---but if the 10\,yr variation in the 800\,MHz pulse profiles of PSR B1937+21 \emph{is} due to the solar cycle, it is difficult to explain how they could display such a strong response when those of the other eight pulsars, whose lines of sight cross much closer to the Sun, do not. 

As discussed above, we observe a low-level ringing in some PCs for this pulsar, visible to some extent in all five PCs in Figure \ref{fig:pca}.5 in the time range MJD 56000--57000. In Figure \ref{fig:lombscargle}, we show Lomb--Scargle periodograms \citep{Lomb, Scargle} of this pulsar's 800\,MHz profiles; a peak appears at $\sim$1.6\,yr. The peak is absent from similar periodograms calculated solely for GUPPI-on data and for 1400\,MHz data, so we interpret it as corresponding to the periodicity of the ringing. Accordingly, we do not attribute the ringing to a solar system effect, either.

\subsubsection{Extrasolar Effects}\label{sec:extrasolar}

\begin{figure}
    \centering
    \includegraphics[width=\columnwidth]{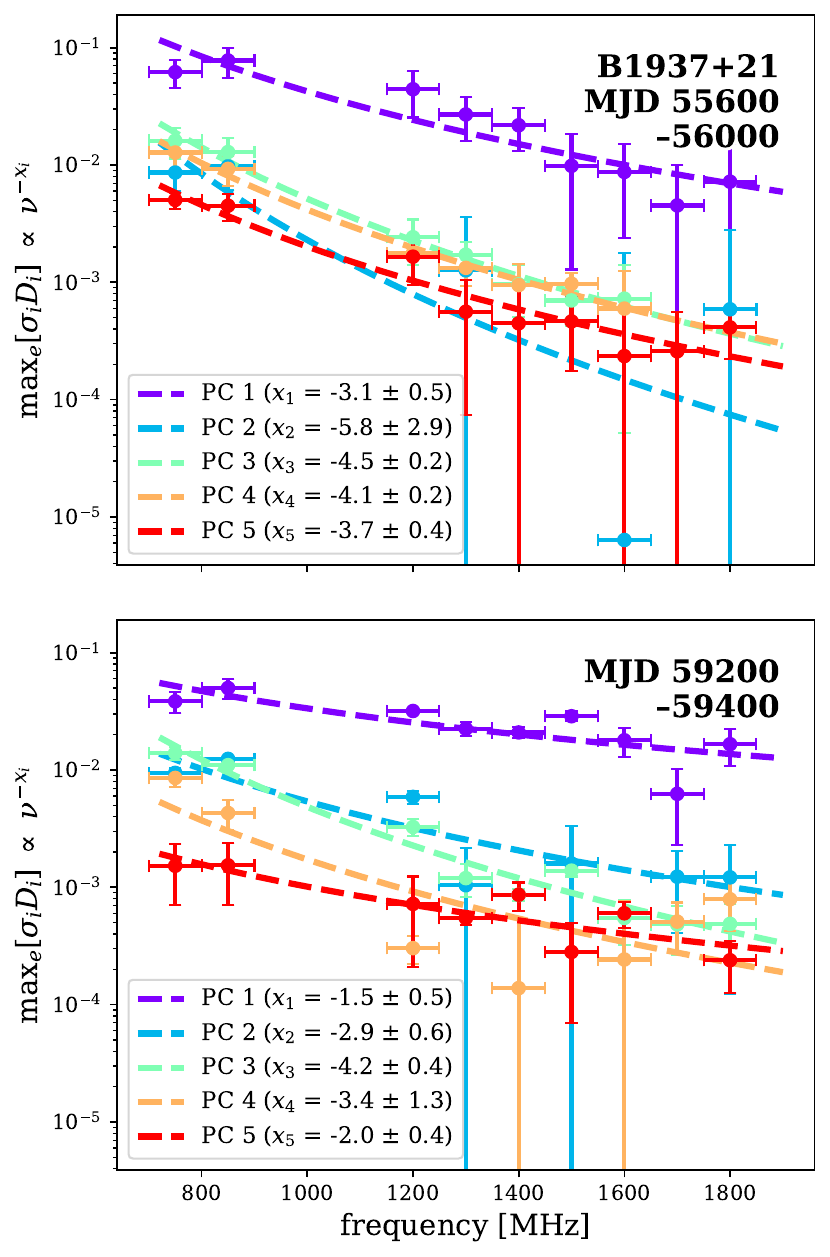}
    \caption{
    Spectral dependence of the (top panel) 2011 and (bottom panel) 2021 pulse shape change events in each of the first five PCs of PSR\,B1937+21. The $y$-axes give the amplitude of each event at each frequency, calculated by fitting a parabola to the points at that frequency in the respective time range. Vertical error bars are $\pm1\sigma$; horizontal error bars are the subband width in frequency. The subscript $e$ in the $y$-axis labels denotes that the maximum is calculated only over the time range of the event. Sometimes the amplitude changes sign at higher frequencies; in these cases, the amplitude is excluded from this plot and not used to calculate the power-law fit.}
    \label{fig:B1937+21chrom}
\end{figure}

We conclude, in agreement with \citet{Brook18} and \citet{Martsen26}, that the long-term variability we see around 800\,MHz and to a lesser extent in the lower $L$ band is likely to have an extrasolar source. \citet{Brook18} show that the expected scattering timescale for PSR\,B1937+21 at 800\,MHz is consistent with the observed pulse shape variation. However, they do not rule out the possibility that the variation is caused by phenomena intrinsic to the pulsar.

We disfavor the intrinsic variability hypothesis because intrinsic changes, including at least three of the four the previously known events analyzed in this dataset, tend to be relatively broadband and decay slowly with frequency. 
Pulse broadening due to scattering, on the other hand, is strongly chromatic, with the scattering timescale\footnote{The subscript ``d'' denotes diffraction, though refraction events are also possible with similar chromatic index. $\tau_{\rm d}$ is not to be confused with the FRED decay timescale, denoted $\tau$, $\tau_0$, or $\tau_j$ in other sections of this paper.} 
$\tau_{\rm d}$ scaling with $\nu^{-x_{\rm d}}$, where $x_{\rm d} = 4.4$ in the limit of small Kolmogorov inner scale and 4 for larger inner scales \citep{CL91}. The scattering timescale does not correspond perfectly to any one PC, but we do observe strong chromaticity in all PCs. In Figure \ref{fig:B1937+21chrom}, we show the amplitude over frequency of the two events highlighted in Figure \ref{fig:B1937+21:regions}. In all PCs for the first event, and in most PCs for the second, the chromaticity is more consistent with scattering than with an intrinsic pulse shape change. (Compare the chromaticity of the 2016 and 2021 events in PSR\,J1713+0747 as shown in Figure \ref{fig:specdep}, for example.)

However, some PCs, especially PC 1, show flatter chromaticity in the 2021 event in PSR\,B1937+21. Further, the pulse shape at 800\,MHz sometimes varies rapidly in a manner consistent with FRED-like events interpreted as magnetospheric disturbances in other pulsars; one such occurrence was the 2021 event, which had a $\zeta$ score of 0.17 (see Figure \ref{fig:othercandidates}). On these grounds, we do not dismiss the intrinsic variability hypothesis entirely.

If the variability is due to scattering, an open question is as to the scattering's source: unlike for PSR\,J1643$-$1224, there is no known H{\sc ii} region along the line of sight to PSR\,B1937+21 \citep{Ocker24}. It is also unclear what could cause two scattering events with such similar effects on pulse shape $\sim$10\,yr apart. 

\citet{Shannon13} describe a circumpulsar asteroid belt as an interpretation for timing variations seen in this pulsar over 26\,yr of TOAs between 1984 November and 2010 June. Included in their analysis are 840\,MHz TOAs from the Westerbork Synthesis Radio Telescope as far back as 2000 January, which includes the time range expected to contain the previous instance of this event if it recurs every 10\,yr. Analysis of the Westerbork observations is beyond the scope of this work, but whether or not the event was observed to occur in $\sim$2001, such analysis would provide valuable clues as to the source of this pulsar's scattering variability.

Future analysis is complicated by the fact that NANOGrav has not regularly observed this pulsar with the GBT's 800\,MHz receiver since 2021.5. Continued observations below 1\,GHz, such as with the 800\,MHz receiver or the new UWBR, will also help characterize the variability whether or not the event occurs again in $\sim$2031. 

\subsection{PSR J1600$-$3053}\label{sec:J1600-3053}

We find three candidate events in PSR\,J1600$-$3053, of which one has a $\zeta$ score of 0.20, lower than two known FRED events. We show the snippet of the $D_2$ time series that resulted in this $\zeta$ score in Figure \ref{fig:othercandidates}. 

This one of two novel candidates with $\zeta_\text{min} \leq 0.25$ that we cannot immediately attribute to an astrophysical source. The other two novel candidates, both in PSR\,B1937+21, clearly correspond to scattering variability of the type discussed by \citet{Brook18} and by us in Section \ref{sec:B1937+21}. The candidate in PSR\,J1600$-$3053, by contrast, is low-S/N and, occurring at an epoch when the GASP back end was in use, can only have been detected in a maximum of two of our sub-averaged frequency channels.

Epochs for this pulsar from much of the time range represented in the right panel of Figure \ref{fig:othercandidates}, especially late 2008, are excised from NANOGrav's pulsar timing pipeline \citep{NG15ObsTiming}. TOAs from the epoch MJD 54614 (i.e., the epoch of the candidate event) and the time range MJD 54765--54827 were excised from both narrowband and wideband timing analysis due to having anomalous DMX values. Since we perform an initial frequency-dependent alignment before subaveraging with PyPulse \citep{PyPulse}, we correct for most epoch-to-epoch DM variation, so it is possible that the anomaly recorded as a DMX event was in fact due to pulse shape variability.

\subsection{PSR J0030+0451}\label{sec:J0030+0451}

PSR\,J0030+0451 is, alongside PSR\,B1937+21, one of the two pulsars in our sample to have a significant interpulse. It has the most observation epochs of any Arecibo pulsar in our sample, contributing to a combined five candidate events between the main pulse and interpulse, including one with $\zeta_\text{min} = 0.11$ (see Figure \ref{fig:othercandidates}).

Like the novel candidate in PSR\,J1600$-$3053, this candidate is low-S/N and we cannot immediately attribute it to an astrophysical source. The low $\zeta_\text{min}$ score is primarily due to the 1400\,MHz subband of PC 3 profiles; while the candidate is detected in four subbands, the adherence to the expected values of $\ell$ and $\alpha$ for a FRED-like event is strongest at 1400\,MHz (see Figure \ref{fig:6pcs_params}.8). A candidate event with $\zeta_\text{min} = 0.32$ is also detected within $\sim$50\,d of this event in the interpulse of this pulsar.

Visual inspection reveals hints of chromaticity in the first PC for both the main pulse and the interpulse similar to what is seen when dot products are taken between PCs and pulse profiles rather than pulse profile residuals, suggesting that updated pulse portraits may be necessary to remove all the frequency-dependent pulse shape for this pulsar. However, since we apply our event search method on a per-frequency basis, the chromaticity in the time series does not affect our findings.

\subsection{PSR J1022+1001}\label{sec:J1022+1001}

\citet{Fiore25} report pulse profile variability on short timescales in PSR J1022+1001, even among subintegrations within a single observing epoch. While we do not address subepoch timescales in this paper, we do observe integrated pulse shape variation from epoch to epoch. The variation appears as single-epoch outliers, similar to those attributed in other pulsars to instrumental noise. 

PSR\,J1022+1001 is the only pulsar in our sample in which we find no candidate events at any $\zeta$, which we attribute to a poor estimate for the noise floor of the CCF. As we discuss in Section \ref{sec:ranking}, FRED--DERF extremum pairs with S/N $<$ 8 in the FRED CCF are discarded, with the noise calculated from the contiguous 1000\,d interval that minimizes the standard deviation. Since PSR\,J1022+1001 has been observed by NANOGrav for such a short time range, and much of that time is excluded from analysis due to the malfunctioning local oscillator at Arecibo, it is difficult to obtain a sensible noise baseline for this pulsar. Of the pulsars in our sample, this observational constraint also affects PSR\,J1903+0327 and PSR\,J2234+0611. More epochs might help define a baseline for PSR\,J1903+0327, which displays strong evidence for refractive scintillation and may have stable periods that can be sensibly considered fiducial, but the time series noise for PSR\,J1022+1001 and PSR\,J2234+0611 appears to be wide-sense stationary on a $\sim$1000\,d timescale, so it is not at all clear that more observations would be useful for this calculation. Rather, our event-finding method, which targets anomalous pulse shape changes, simply may not be suitable for identifying epoch-to-epoch pulse shape variation in pulsars like PSR\,J1022+1001. However, the variation is readily apparent to the eye from the $D_i$ time series. Future work, including \citet{Martsen26}, will address other methods for characterizing variability in these time series.

\subsection{PSR J1903+0327}\label{sec:J1903+0327}

PSR\,J1903+0327 is known to exhibit strong scattering-induced pulse shape variability on a characteristic timescale of $\sim$100\,d \citep{Geiger25}. Its pulse profiles have, by a large margin, the largest variance of any pulsar in our sample (see the right-hand panels of the Figure \ref{fig:comps} figure set). In the $D_i$ time series, this scattering-induced variability is obvious to the eye.

In PSR\,B1937+21 and PSR\,J1643$-$1224, the scattering variability affects pulse profiles most strongly below 1\,GHz. Figure \ref{fig:B1937+21chrom} shows that the dropoff with increasing frequency is rapid. For PSR\,J1903+0327, we analyze only pulse profiles from above 1\,GHz and still see a high degree of variability.

As discussed in Section \ref{sec:J1022+1001}, the high scattering variability of this pulsar combined with the relatively short interval of observations hinders our ability to find one-off FRED-like events, if indeed any exist. Our method finds three candidates, with the best at a relatively high $\zeta$ score of 0.51.

\section{Conclusion}\label{sec:conc}

FRED-like pulse shape change events have been observed to occur in at least two pulsars in the NANOGrav PTA. One of them, PSR\,J1713+0747, has undergone three events and would make an excellent timing pulsar if not for the effect of such drastic pulse shape changes on precision timing. In pursuit of mitigating the impact of potential low-level events on timing noise, we present an event search method and apply it to nine pulsars in the NANOGrav PTA.

Our method leverages PCA and matched filtering. With PCA, we construct time series of PC dot products with pulse profile residuals (Figure \ref{fig:pca}), which provide a tool for visualizing pulse shape variability. These time series can be studied in a variety of ways, including to characterize long-term variability \citep{Martsen26}, but we focus this work on a search for sparse FRED-like events.
We employ a dual matched filter with FRED and DERF templates and use statistical parameters from their CCF responses to implement a ranking metric $\zeta$, given in Equation \ref{eq:zeta}.

We recover all four known FRED-like pulse shape change events, shown in time series in Figure \ref{fig:knowncandidates}. In a timing analysis, sparsely occurring events like these can be dealt with in two ways: by excising the events from the dataset before conducting the analysis or by model fitting the events as part of the analysis. In Section \ref{sec:J1713+0747}, we calculate exclusion times for the three events in PSR\,J1713+0747. Collectively, excising the events would create timing gaps totaling $\sim$6\,yr. Model fitting an event as part of a timing analysis is a better approach if the event shape can be suitably parameterized, but this requires further study.

We also detect four new candidate events with $\zeta$ scores better than the worst of the known events, shown in their respective time series in Figure \ref{fig:othercandidates}. We attribute two of them to scattering variability in PSR\,B1937+21. The other two candidates, in PSR\,J0030+0451 and PSR\,J1600$-$3053, are difficult to characterize because of their low S/N in the $D_i$ time series. However, within $\sim$50\,d of the PSR\,J0030+0451 event, a second candidate is detected at moderately low $\zeta$ score in the interpulse of that pulsar, suggesting that the shape change corresponding to the primary candidate occurs at some level over both pulses. Meanwhile, the time range of the PSR\,J1600$-$3053 event is excised from the NANOGrav timing pipeline on the basis of DMX.

One of the novel candidates in PSR\,B1937+21 corresponds to the recurrence of a pulse shape change event (albeit not a FRED-like one) previously observed $\sim$10\,yr before in the same pulsar. We rule out instrumental error and solar systematics as causes and tentatively attribute both events to scattering variability, though it is unclear what scattering process would produce two nearly identical pulse shape change events $\sim$10\,yr apart. We highlight that the events occur most strongly at $\sim$800\,MHz but that NANOGrav stopped observing PSR\,B1937+21 at 800\,MHz with the GBT in 2021.5. The resumption of 800\,MHz observations, such as with the GBT's new UWBR \citep{UWBR20, UWBR23}, would help determine whether these events are periodic. Analysis of archival observations from before the NANOGrav PTA may also do this, but is outside the scope of this work.

Our analysis was conducted over 100\,MHz subaveraged frequency channels, so we retain a high-level sensitivity to the chromaticity of events. In the discovery paper of the PSR\,J1643$-$1224 event, \citet{Shannon16} observe the event at 1500 and 3000\,MHz but not at 600\,MHz; we further constrain its frequency dependence with a detection at 1400\,MHz and a nondetection at 1300\,MHz. \citet{Jennings24} discuss the frequency dependence of the 2021 PSR\,J1713+0747 event over 25\,MHz channels for TOA residuals and receiver-to-receiver for PCA dot products; we extend their PCA analysis to finer frequency scales and show the chromaticity of the shape change in the second PC in Figure \ref{fig:specdep}. Future pulsar observations over wide frequency ranges, such as with the GBT UWBR, may provide an immediate way to mitigate pulse shape change events by enabling narrowband timing in bands that show no significant variation.

\section*{Acknowledgments}

This work was supported by National Science Foundation (NSF) Physics Frontiers Center award Nos. 1430284 and 2020265.
P.R.B.\ is supported by the Science and Technology Facilities Council, grant number ST/W000946/1.
H.T.C.\ acknowledges funding from the U.S. Naval Research Laboratory.
Pulsar research at UBC is supported by an NSERC Discovery Grant and by CIFAR.
K.C.\ is supported by a UBC Four Year Fellowship (6456).
M.E.D.\ acknowledges support from the Naval Research Laboratory by NASA under contract S-15633Y.
T.D.\ and M.T.L.\ received support by an NSF Astronomy and Astrophysics Grant (AAG) award number 2009468 during this work.
E.C.F.\ is supported by NASA under award number 80GSFC24M0006.
D.C.G.\ is supported by NSF Astronomy and Astrophysics Grant (AAG) award \#2406919.
D.R.L.\ and M.A.M.\ are supported by NSF \#1458952.
M.A.M.\ is supported by NSF \#2009425.
The Dunlap Institute is funded by an endowment established by the David Dunlap family and the University of Toronto.
T.T.P.\ acknowledges support from the Extragalactic Astrophysics Research Group at E\"{o}tv\"{o}s Lor\'{a}nd University, funded by the E\"{o}tv\"{o}s Lor\'{a}nd Research Network (ELKH), which was used during the development of this research.
H.A.R.\ is supported by NSF Partnerships for Research and Education in Physics (PREP) award No.\ 2216793.
S.M.R.\ and I.H.S.\ are CIFAR Fellows.
Portions of this work performed at NRL were supported by ONR 6.1 basic research funding.

\appendix

\section{Full Candidate Event List}\label{app:list}

In Table \ref{tab:allcandidates}, we provide a full list of candidate events found by our search method in our nine-pulsar sample. A horizontal line separates the entries in Table \ref{tab:candidates} from the rest; the cutoff is placed immediately after the last recovery of a known FRED-like event. The entries from the $\zeta_\text{min}$ column are shown per pulsar in Figure \ref{fig:zetadist}. 

\begin{center}
\begin{table*}
\begin{threeparttable}
    \caption{All Candidate Events.}
    \begin{tabular}{l l c | c c c c c}
        \hline
        \hline
        Pulsar  & MJD & $\zeta_\text{min}$ & $\zeta_1$ & $\zeta_2$ & $\zeta_3$ & $\zeta_4$ & $\zeta_5$\\
        \hline
J1713+0747&59330&0.05& 0.05& 0.08& 0.13& 0.07& 0.10\\
J1713+0747&57520&0.08& --- & 0.08& 3.00& --- & 0.44\\
J0030+0451&57290&0.11& 0.39& --- & 0.11& 0.49& --- \\
B1937+21&57040&0.16& 0.85& 0.16& 2.72& --- & --- \\
B1937+21&59190&0.17& 0.54& 0.27& 0.17& 0.34& 0.36\\
J1600$-$3053&54610&0.20& 0.60& --- &0.2& --- & --- \\
J1643$-$1224&57100&0.24& 0.55& 0.24& 5.79& 0.88& --- \\
J1713+0747&54780&0.25& 0.75& 0.25& 0.52& --- & --- \\
\hline
J1643$-$1224&55340&0.30& 0.30& 0.36& --- & 1.50& --- \\
J0030+0451i&57340&0.32&0.32& 0.43& --- & --- & --- \\
J1600$-$3053&57670&0.40& 2.78& 0.40&0.64& 4.23& --- \\
B1937+21&55640&0.44& 0.44& 1.30& 0.49& 0.71& 1.02\\
B1937+21i&55620&0.47& 0.47& 1.57& --- & 6.90& --- \\
B1937+21i&58760&0.49& --- & --- & --- & 0.49& --- \\
J1903+0327&57520&0.51& --- & --- & --- & --- & 0.51\\
J1643$-$1224&53840&0.51& --- & --- & 0.51& --- & --- \\
J1909$-$3744&55540&0.62& 0.62& 0.86& --- & --- & 2.86\\
J1909$-$3744&53980&0.64& --- & 0.64& --- & --- & --- \\
J1643$-$1224&57550&0.65& 0.65& 1.05& 4.67& 1.44& --- \\
J1600$-$3053&56310&0.68& --- & 2.64&0.68& --- & --- \\
B1937+21i&55860&0.68& --- & --- & 0.68& 3.24& --- \\
J1600$-$3053&55340&0.70& 3.22& 0.90&0.70& --- & --- \\
J1909$-$3744&57170&0.70& --- & 0.70& --- & --- & --- \\
J1713+0747&55220&0.72& --- & 0.72& --- & --- & --- \\
J1903+0327&56970&0.79& --- & --- & 0.79& --- & --- \\
J1909$-$3744&57740&0.90& --- & --- & 0.90& --- & 0.99\\
B1937+21&58730&0.90& --- & --- & --- & 0.90& 1.26\\
B1937+21&54980&0.92& 0.92& --- & --- & --- & --- \\
J1909$-$3744&56920&0.98& 0.98& 34.98& 2.67& --- & 2.24\\
J1643$-$1224&55670&1.01& 1.01& --- & 3.45& 11.83& 5.85\\
J1713+0747&56590&1.08& 1.08& --- & 1.52& 2.91& --- \\
J1903+0327&58710&1.14& --- & --- & --- & 1.14& --- \\
J1643$-$1224&56130&1.18& --- & --- & 1.18& --- & --- \\
B1937+21&53300&1.19& --- & 1.19& --- & --- & --- \\
J0030+0451i&56240&1.27&1.27& --- & --- & --- & --- \\
J1909$-$3744&57470&1.31& --- & 1.31& --- & --- & --- \\
J1713+0747&59800&1.32& --- & --- & --- & 1.32& --- \\
J0030+0451&56240&1.41& --- & --- & --- & 1.41& --- \\
J1909$-$3744&56240&1.50& 1.50& --- & 62.55& --- & --- \\
J1713+0747&57160&1.73& 1.73& --- & --- & --- & --- \\
J0030+0451&56870&2.77& --- & --- & --- & 2.77& --- \\
B1937+21i&57890&2.81& --- & 2.81& --- & --- & --- \\
B1937+21&58130&2.87& --- & 2.87& --- & --- & --- \\
J1713+0747&58840&3.85& --- & --- & --- & 3.85& --- \\
J2234+0611&57890&4.66& --- &4.66& --- & --- & --- \\
J1643$-$1224&55000&9.83& --- & --- & --- & 9.83& --- \\
B1937+21i&54370&10.11& --- & --- & --- & 10.11& --- \\
        \hline
    \end{tabular}
    \begin{tablenotes}
        \footnotesize
        \item[a] In Table \ref{tab:candidates}, these epochs are given as reported in the events' respective discovery papers. In this table, they are given as recorded by our search method. In all cases, the difference is $\lesssim$30\,d, the approximate cadence of NANOGrav observations for a given pulsar.
    \end{tablenotes}
    \label{tab:allcandidates}
\end{threeparttable}
\end{table*}
\end{center}

\section{Nonuniform Time Sampling}\label{app:nonuniform}

\begin{figure}
    \centering
    \includegraphics[width=0.45\linewidth]{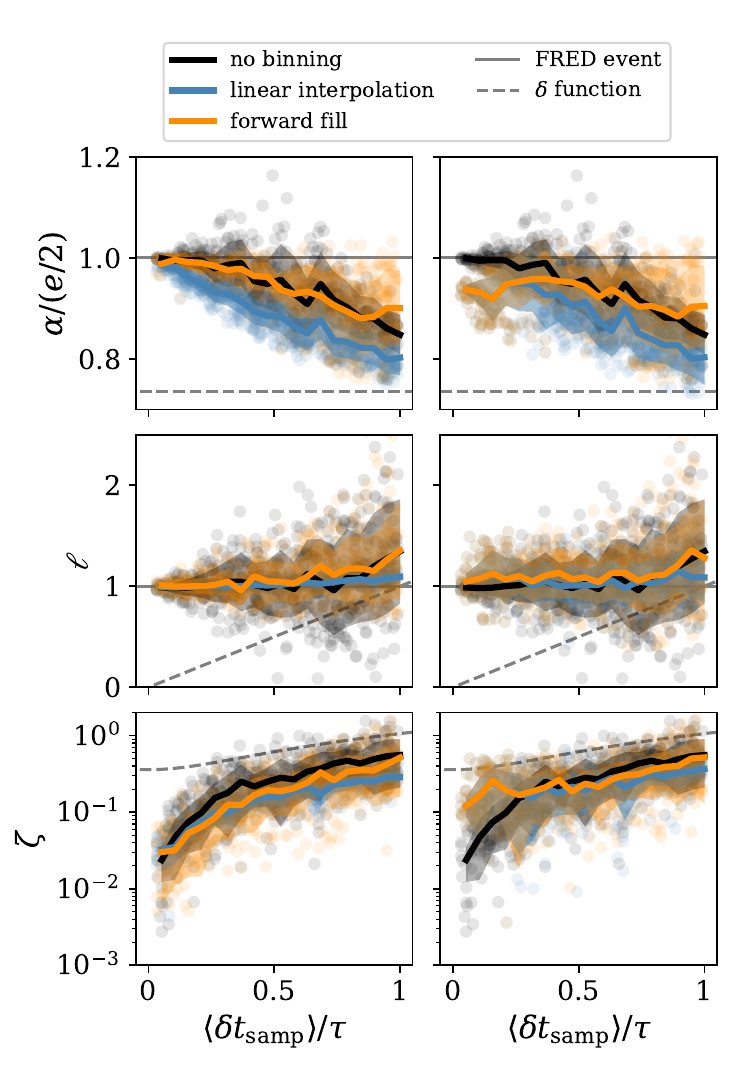}
    \caption{Effect of nonuniform time sampling on FRED--DERF amplitude ratio $\alpha$, dimensionless lag $\ell$, and $\zeta$ score (Equation \ref{eq:zeta}). Here we show 500 trials of the same FRED event sampled at different average rates $\langle \delta t_{\rm samp}\rangle$, binned according to various prescriptions, then matched with a FRED and a DERF template of the same decay time $\tau$. Each sample was allowed to occur earlier or later than the fiducial time by up to $\langle\delta t_{\rm samp}\rangle/2$. In the left-column panels, we show a bin width of 0.033 times the decay time of the pulse; in the right-column panels, the bin width is 0.300 times the decay time. The black curves give the no-binning case (used in our search) and are identical across columns. The colored curves give different prescriptions for binning in the case of an unfilled bin: either taking the average of adjacent values (blue) or carrying over the previous value (orange). 
    }
    \label{fig:unevensamp}
\end{figure}

In Section \ref{sec:methods}, we describe a framework for searching for FRED-like events that involves calculating the CCF of the $D_i$ time series with a series of FRED templates with decay times between 50 and 400. In that section, we assume uniform time sampling to calculate detection parameters $\ell$ and $\alpha$ that help us filter CCF extrema for event candidates, then rank the candidates using the $\zeta$ score (Equation \ref{eq:zeta}). We show a toy model demonstrating $\ell$ and $\alpha$ in this idealized case in Figure \ref{fig:matchfiltertoy}. 

However, NANOGrav MSP observations are nonuniform. The observing cadence for most pulsars in the 15\,yr dataset was on average monthly ($\sim$3 weeks at Arecibo and $\sim$4 weeks with the GBT), except for a few pulsars that were observed weekly \citep{NG15ObsTiming}. In our sample, PSR\,J0030+0451, PSR\,J1713+0747, and PSR\,J1909$-$3744 fall into the latter category \citep{NG12}. These cadences could shift by up to several days depending on the observing schedule, and additional gaps are created in our analysis by the rejection of profiles with S/N below the cutoffs given in Table \ref{tab:snrcuts}.

For the method discussed in Section 3, we calculate the CCF of the $D_i$ time series with a template by iteratively evaluating the template at the observation epochs of the time series, taking the dot product of the two vectors, and then shifting the template by a delay time $\delta t_{\rm temp} \ll \langle \delta t_{\rm samp}\rangle$, where, as in Section \ref{sec:methods}, $\langle \delta t_{\rm samp}\rangle$ is the average sampling cadence of the time series. In this way, we build a well-sampled CCF where any discontinuities are the result of the observation schedule and/or low-S/N rejection. Example CCFs are shown in Figures \ref{fig:matchfilterex} and \ref{fig:matchfilterheatmap}.

An alternative we tested was to bin the $D_i$ time series to a uniform sampling rate. We considered bin widths $\delta t_{\rm bin}$ less than and greater than $\langle \delta t_{\rm samp}\rangle$ and applied them to the toy model used for Figure \ref{fig:matchfiltertoy}: a FRED event with a decay time $\tau = 300$ in simulated data of length 2048. We sampled the data with randomly drawn average cadences between 10 and 300, then permitted each sample to occur early or late by up to half the average cadence.

In the case that $\delta t_{\rm bin}$ is less than any interval between consecutive observations $\delta t_{\rm samp}$, one or more bins may be unfilled. We tested two methods for filling them: a forward fill, for which we propagated the previous bin value to any empty bins; and a linear interpolation, in which we assigned any empty bins the mean value of the two adjacent bins.

Figure \ref{fig:unevensamp} shows the recovery of the event using $\alpha$, $\ell$, and $\zeta$ over 500 trials. For each trial, the event was sampled to a different average cadence $\langle \delta t_{\rm samp}\rangle$ between 0 and 1 times the event decay time $\tau$. (For our search, which prioritizes a search range of 50--400\,d for $\tau_j$, we are nearly always below $0.5\langle \delta t_{\rm samp}\rangle/\tau$.) Each individual sample was allowed to vary by up to $\langle \delta t_{\rm samp}\rangle / 2$. The black curves show the recovery of the event without binning, as is done for our search. The blue curves represent binning with the linear interpolation method for filling empty bins and the orange curves represent binning with the forward fill method. In the left column, $\delta t_{\rm bin} = 0.033 \tau$; in the right column, $\delta t_{\rm bin} = 0.300 \tau$. The shaded region surrounding each curve gives the standard deviation of the trials. In the finely binned case (left column), the curves agree to within this uncertainty, but in the coarsely binned case, where smoothing of the peak becomes significant when $\delta t_{\rm samp} < \delta t_{\rm bin}$, the recovered $\alpha$ and $\zeta$ values diverge and become worse when binned. Future work with this method could further investigate the optimal $\delta t_{\rm bin}$ in the fine-binning regime, but we avoid making prescriptions about binning for this work by using the unbinned case (the black curve) for our search.

\bibliographystyle{aasjournal}
\bibliography{main.bib}

\end{document}